\documentclass[pdflatex,sn-nature]{sn-jnl}

\usepackage{graphicx}%
\usepackage{multirow}%
\usepackage{amsmath,amssymb,amsfonts}%
\usepackage{amsthm}%
\usepackage{mathrsfs}%
\usepackage[title]{appendix}%
\usepackage{xcolor}%
\usepackage{textcomp}%
\usepackage{manyfoot}%
\usepackage{booktabs}%
\usepackage{algorithm}%
\usepackage{algorithmicx}%
 \usepackage{soul}%
\usepackage{algpseudocode}%
\usepackage{listings}%
\usepackage{lineno}%
\usepackage{xr-hyper}

\theoremstyle{thmstyleone}%
\theoremstyle{thmstyletwo}%

\theoremstyle{thmstylethree}%

\begin{document}

\title[Article Title]{Antireflection by design in bilayer metasurfaces}


\author*[1,2]{\fnm{Jaewon} \sur{Oh}} \email{ohj7@corning.com}
\equalcont{These authors contributed equally to this work.}

\author[1]{\fnm{Davide} \sur{Cassara}}
\equalcont{These authors contributed equally to this work.}

\author[1]{\fnm{Alfonso} \sur{Palmieri}}
\equalcont{These authors contributed equally to this work.}

\author[1]{\fnm{Lorenzo} \sur{Piatti}}

\author[2]{\fnm{Janderson} \sur{Rocha Rodrigues}}

\author[1,3]{\fnm{Ahmed H.} \sur{Dorrah}}

\author[2]{\fnm{Paulo} \sur{Dainese}}

\author*[1]{\fnm{Federico} \sur{Capasso}}\email{capasso@seas.harvard.edu}

\affil[1]{\orgdiv{Harvard John A. Paulson School of Engineering and Applied Sciences}, \orgname{Harvard University}, \orgaddress{\city{Cambridge}, \postcode{02138}, \state{Massachusetts}, \country{USA}}}

\affil[2]{\orgdiv{Corning Research and Development Corporation}, \orgname{Corning Inc.}, \orgaddress{\city{Painted Post}, \postcode{14870}, \state{New York}, \country{USA}}}

\affil[3]{\orgdiv{Department of Applied Physics and Science Education}, \orgname{Eindhoven University of Technology}, \orgaddress{\city{Eindhoven}, \postcode{5600}, \country{The Netherlands}}}


\abstract{
Antireflection coatings are ubiquitous in optical systems, where they maximize transmission and suppress undesirable reflections by impedance-matching uniform interfaces. Extending this principle to metasurfaces, however, is fundamentally more challenging because wavefront control requires a library of geometrically distinct meta-atoms, each locally imposing a prescribed phase that is tethered to its transmittance. Here, we show that vertical integration resolves this constraint by allowing bilayer meta-atoms to operate simultaneously as a phase shifter and an impedance-matching stack. Using an effective thin-film model, we derive a design rule that links the effective indices of two independently patterned layers and identifies antireflective bilayer libraries with full {\unboldmath$0$--$2\pi$} transmission-phase coverage. We realize this concept in a free-standing TiO$_2$/TiO$_2$ metalens operating at 1310~nm, which suppresses reflectance below that of bare glass while preserving diffraction-limited focusing. These results establish bilayer metasurfaces as a framework for co-engineering optical impedance and wavefront response at the meta-atom level.
}



\maketitle

Metasurfaces shape optical wavefronts using subwavelength scatterers that impart prescribed local responses on incident light \cite{yu2014}. This principle has enabled compact flat-optical components for imaging \cite{khorasaninejad2016, arbabi2023, park2024}, holography \cite{huang2018metasurface, ren2019}, beam shaping, structured-light generation \cite{dorrah2022tunable, yessenov2025ultrafast}, and fiber-based communications \cite{oh2022adjoint, oh2023metasurfaces}, among others. In these applications, optical throughput is not merely a device-level figure of merit. Reflection losses reduce the photon budget, generate parasitic back-reflections, produce stray light and ghost images, and complicate integration with surrounding optical systems \cite{shi2026exploring, ding2024breaking}. Dielectric metasurfaces are especially attractive because they combine low absorption, high transmission and nearly arbitrary phase control \cite{kamali2018, khorasaninejad2016polarization}. Many efforts to improve metasurface efficiency have focused on controlling the transmitted field more accurately, including topology optimization \cite{sell2017, phan2019, mansouree2019metasurface} and shape optimization \cite{dainese2024shape}. These approaches have substantially advanced wavefront fidelity and diffraction efficiency, but they generally do not explicitly eliminate reflection losses at the air--metasurface and metasurface--substrate interfaces.

\begin{figure}[h]
\centering
\includegraphics[width=1.00\linewidth]{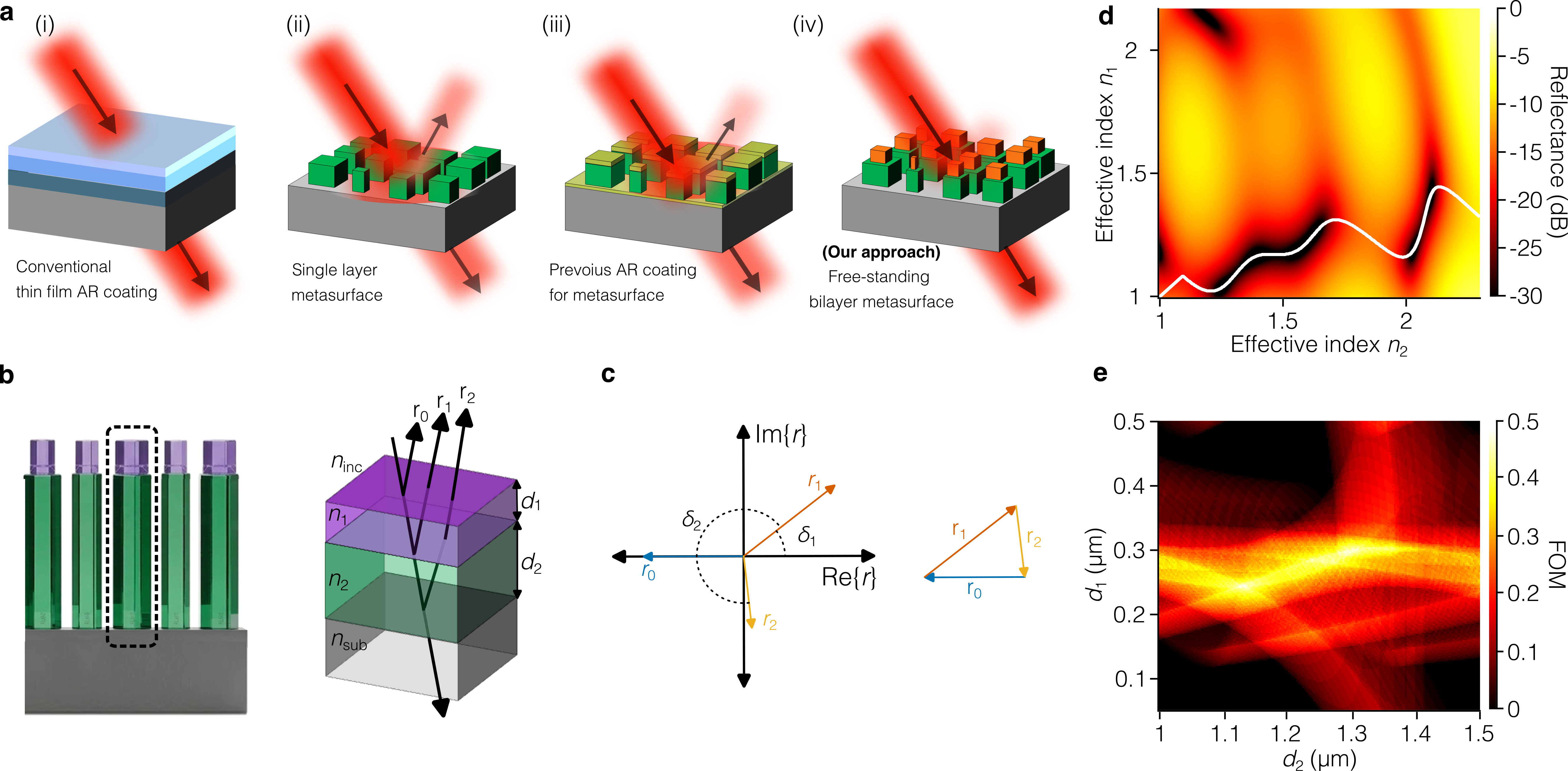}
\caption{\textbf{Thin-film model of antireflective bilayer meta-atoms.}
\textbf{a,} Conceptual comparison of antireflection strategies. A conventional thin-film AR coating on a flat interface yields negligible reflection ($R \approx 0\%$). A single-layer metasurface exhibits residual reflection ($R \approx 6\%$), and adding a conventional AR coating on top suppresses the reflection only partially ($R \approx 4\%$). By contrast, the free-standing bilayer metasurface approach introduced here reduces the reflection to $2\%$.
\textbf{b,} Cross-sectional schematic of the bilayer metasurface and corresponding effective thin-film model. A representative bilayer meta-atom, highlighted by the dashed box, is approximated as two homogeneous films with effective refractive indices $n_1$ and $n_2$ and thicknesses $d_1$ and $d_2$, placed between the incident medium $n_{\mathrm{inc}}$ and the substrate $n_{\mathrm{sub}}$. The dominant reflected contributions, $r_0$, $r_1$ and $r_2$, arise from the three interfaces.
\textbf{c,} Geometric interpretation of the reflected fields in the complex plane. Zero reflection is obtained when the three reflected components interfere destructively and form a closed triangle.
\textbf{d,} Calculated reflectance of the effective thin-film stack at normal incidence for $\lambda = 1310$~nm, plotted as a function of the effective indices $n_1$ and $n_2$ for fixed film thicknesses. The white solid curve marks the locus of index pairs that minimize the reflectance for each value of $n_2$.
\textbf{e,} Figure of merit (FOM) used to optimize the thicknesses $d_1$ and $d_2$.}
\label{fig:Vector}
\end{figure}

A complementary route is to reduce parasitic reflection through antireflection design. In conventional optics, reflection at an interface can be suppressed by impedance matching through a thin-film coating~\cite{macleod2010thin} (Fig.~\ref{fig:Vector}a(i)). Extending this concept to metasurfaces, however, is fundamentally non-trivial. The difficulty is not simply that metasurfaces reflect light; it is that they have a spatially varying reflectivity. Each building block, or meta-atom, composing the metasurface is obtained from a library of nanopillar geometries with fixed height but varying lateral dimensions. A meta-atom approximates an effective refractive index which determines the transmission phase, amplitude and therefore its reflection (Fig.~\ref{fig:Vector}a(ii)). A uniform antireflection (AR) coating placed on top of such a library can satisfy the impedance-matching condition only approximately, and only for a subset of meta-atoms ~\cite{miyata2024anti, koksal2024antireflective,ryu2022high, hu2024design} (Fig.~\ref{fig:Vector}a(iii)). Furthermore, a top coating addresses only the air--metasurface interface, leaving the metasurface--substrate interface, a persistent source of reflection loss, untreated unless a separate bottom coating is also applied.

Recent advances in bilayer metasurface fabrication provide a route to address this library-level constraint by embedding antireflection directly into the meta-atom architecture rather than adding it after fabrication \cite{dorrah2025free, dorrah2025pushing}. Here we use two vertically stacked nanopillar layers with independently controlled lateral geometries (Fig.~\ref{fig:Vector}a(iv)) and model each bilayer meta-atom as an effective four-layer thin-film stack consisting of the incident medium, the two nanopillar layers and the substrate. In this picture, vertical integration changes the role of the meta-atom: each element acts not only as a local phase shifter, but also as a local impedance-matching stack. Because the two patterned layers provide independently tailorable effective indices \cite{cassara2025antireflective}, the bilayer architecture supplies the additional degrees of freedom needed to reduce reflection from the full stack while preserving the phase coverage required for wavefront shaping. In other words, the metasurface is not impedance matched after the phase library is designed; impedance matching is built into the library itself.

Using this framework, we design TiO$_2$/TiO$_2$ bilayer metalenses operating at $\lambda_d = 1310$~nm. Full-device simulations show that the bilayer architecture reduces the reflection loss by more than a factor of two, from 5.6\% for a conventional single-layer metalens to 2\%, below the 3.3\% reflection of a bare glass substrate, without compromising wavefront formation: the bilayer and single-layer devices exhibit nearly identical point-spread functions and Strehl ratios. Because more light is funneled through the metasurface at no cost to phase error, the overall focusing efficiency increases. Experimentally, the reduction in parasitic reflection is accompanied by an increase in focusing efficiency from 86\% for the single-layer control to 90\% for the bilayer metalens with the same aperture and focal length. A gain of roughly 4 percentage points in absolute focusing efficiency may appear modest in isolation, but it is consequential in the applications where flat optics face the most competitive pressure, such as cascaded imaging or coupling to photonic integrated circuits.

These results establish vertically integrated metasurfaces as a practical framework for co-engineering optical impedance and wavefront response, shifting antireflection from a post-fabrication coating problem to a metasurface design principle in which the phase library itself is impedance matched.

\section*{Design principle}\label{secdesign}
\subsection*{Bilayer meta-atom as an antireflective thin-film stack}\label{subsecthinfilm}

The design problem is to construct a meta-atom library that is both phase-addressable and antireflective. A single patterned layer with fixed height and varying lateral geometry cannot satisfy both requirements simultaneously across the full $0$--$2\pi$ phase range: meta-atoms at different phase levels present different effective indices to the incident medium, so no single impedance-matching condition can be enforced uniformly across the library.  The bilayer architecture supplies the missing degree of freedom — the upper-layer geometry — which can be tuned independently for each phase level to suppress reflection without disturbing the phase response set by the lower layer.

Figure~\ref{fig:Vector}b illustrates the approximation that connects this library problem to thin-film antireflection theory. When the response of a subwavelength pillar lattice is governed by the Bloch mode, the lattice can be represented by a homogeneous medium with an effective refractive index given by the modal propagation constant~\cite{lalanne2006optical, kikuta1998effective, lalanne1996effective}. A bilayer meta-atom is then modelled as a four-layer stack with refractive indices $n_{\mathrm{inc}}$, $n_1$,
$n_2$ and $n_{\mathrm{sub}}$, where $n_1$ and $n_2$ are the Bloch-mode effective indices of the upper and lower layers, respectively, and $d_1$ and $d_2$ are their physical \cite{song2023transfer}.

This effective-stack description reduces the metasurface antireflection problem to that of a classical bilayer antireflection coating \cite{schuster1949anwendung, cox1961special}. We analyze the reflectance with the vector method~\cite{macleod2010thin}; a transfer-matrix calculation gives a nearly indistinguishable reflectance map (Supplementary Section~S1). At normal incidence, the reflected amplitudes at the three interfaces are
\begin{equation}
    r_0=\frac{n_{\mathrm{inc}}-n_1}{n_{\mathrm{inc}}+n_1}, \qquad
    r_1=\frac{n_1-n_2}{n_1+n_2}, \qquad
    r_2=\frac{n_2-n_{\mathrm{sub}}}{n_2+n_{\mathrm{sub}}},
    \label{eqfrsnel}
\end{equation}
for either TE- or TM-polarized light. The phase thickness accumulated in the $m^{\mathrm{th}}$ film is
\begin{equation}
    \delta_m=\frac{2\pi}{\lambda}n_m d_m,
    \label{eqdelta}
\end{equation}
and the reflectance of the effective bilayer stack is therefore
\begin{equation}
    R(n_1,n_2,d_1,d_2)=\left|r_0+r_1 e^{2 i \delta_1}+r_2 e^{2 i (\delta_1+\delta_2)}\right|^2.
    \label{eqReflectance}
\end{equation}

Equation~\ref{eqReflectance} gives a simple geometric interpretation of antireflection. The three terms correspond to reflected fields from the air--upper-layer, upper--lower-layer and lower-layer--substrate interfaces. In the complex plane, zero reflection is obtained when these three vectors interfere destructively and form a closed triangle (Fig.~\ref{fig:Vector}c). The side lengths are set by the Fresnel amplitudes, while the angles by the layer phase thicknesses.

This picture also shows why antireflection is hard to enforce across a library. For each pair of effective indices, exact triangle closure requires specific phase thicknesses $\delta_1$ and $\delta_2$. A metasurface, however, has fixed layer heights $d_1$ and $d_2$, so the exact cancellation condition cannot be enforced for every meta-atom. The thicknesses must instead be co-optimized so the library as a whole remains weakly reflective.

We apply this design principle to TiO$_2$/TiO$_2$ bilayer meta-atoms on glass, with $n_{\mathrm{sub}}=1.447$ and $n_{\mathrm{inc}}=1$, at a design wavelength of $\lambda=1310$~nm. For a representative thickness pair, $d_1=290$~nm and $d_2=1300$~nm, equation (\ref{eqReflectance}) gives the reflectance map shown in effective index-space in Fig.~\ref{fig:Vector}d. Each point $(n_1,n_2)$ represents one possible bilayer meta-atom.

Fabrication imposes $n_1 \leq n_2$ \cite{palmieri2024dielectric}, and the lower pillar is also substantially taller because the two electron-beam resists have different accessible thickness windows; we therefore assign most of the transmission phase to the lower layer.

The optimal path is obtained by defining a mapping
\begin{equation}
    f(n_2)=n_1,
    \label{eqmapping}
\end{equation}
such that
\begin{equation}
    R_{\mathrm{min}}(n_2,d_1,d_2)=R\!\left(f(n_2),n_2,d_1,d_2\right),
    \label{eqmapRefl}
\end{equation}
where $R_{\mathrm{min}}$ is the minimum reflectance attainable for a given lower-layer effective index and fixed thicknesses $d_1$ and $d_2$. Thus, $f$ assigns to each phase-setting lower layer the upper-layer effective index that best impedance matches the full stack. The resulting minimum-reflectance locus is shown by the white solid curve in Fig.~\ref{fig:Vector}d. Repeating this optimization over all viable thickness pairs gives a family of mappings $f$ and corresponding minimized-reflectance curves. The layer thicknesses $d_1$ and $d_2$ are then chosen to maximize the fraction of the library along this locus that falls below $R = -30$~dB; the optimal pair, $(d_1, d_2) = (290~\mathrm{nm}, 1300~\mathrm{nm})$, is shown by the marker in Fig.~\ref{fig:Vector}e (see Supplementary Section~S2 for more detailed discussion).

The mapping $f(n_2)$ is the core result of the thin-film model. It converts antireflection from a global coating condition into a metasurface library design rule: for each meta-atom, it specifies the upper-layer effective index that minimizes reflection. The upper layer is therefore not a uniform coating, but a phase-level-dependent impedance-matching layer co-designed with the phase-setting lower geometry. Representative solutions in Supplementary Section S2 confirm the geometric origin of the antireflective response.

\subsection*{Antireflective bilayer meta-atom library construction}\label{subseclibrarydesign}

The thin-film model defines the target relation between the two layers. We translate this effective-index rule into physical nanopillar geometries using rigorous coupled-wave analysis (RCWA), implemented with Reticolo \cite{hugonin2021reticolo}, which directly returns the Bloch-mode propagation constants of each periodic structure.

\begin{figure}[h!]
\centering
\includegraphics[width=1.0\linewidth]{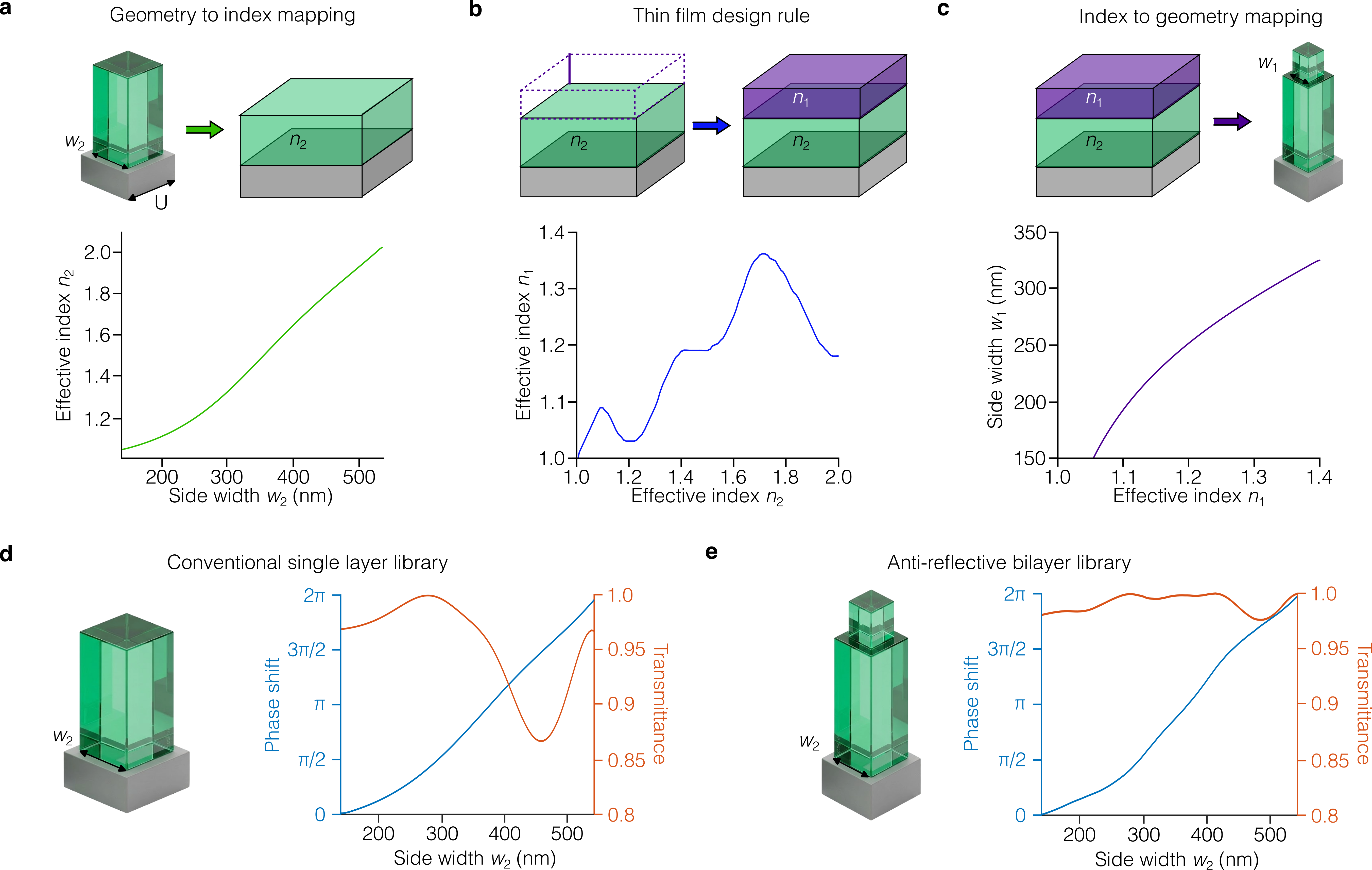}
\caption{\textbf{Library design of antireflective bilayer meta-atoms at 1310~nm.}
\textbf{a,} Effective index of the fundamental Bloch mode in periodic square TiO$_2$ nanopillars as a function of side width $w_2$, for a pitch $U$ of 600~nm.
\textbf{b,} Thin-film mapping that assigns, for each effective index $n_2$ of the lower pillar, the effective index $n_1$ of the upper pillar that minimizes the reflectance of the bilayer meta-atom.
\textbf{c,} Inverse mapping of panel a, used to convert the target effective index of the upper layer back into a physical side width.
\textbf{d,} Simulated phase shift and transmittance of a conventional single-layer TiO$_2$ meta-atom library.
\textbf{e,} Simulated phase shift and transmittance of the antireflective bilayer TiO$_2$ meta-atom library. Both libraries provide full $0$--$2\pi$ phase coverage, but the bilayer design maintains higher transmittance across the full library.}
\label{fig:Library}
\end{figure}

Figure~\ref{fig:Library}a shows the effective index $n_{\mathrm{eff}} = g(w)$ of the fundamental Bloch mode for periodic square TiO$_2$ nanopillars with pitch $U=600$~nm at $\lambda=1310$~nm; a square cross-section was chosen for polarization insensitivity \cite{kikuta1998effective}. Combining $g$ with the thin-film mapping $f$ (Fig.~\ref{fig:Library}b) and the inverse $g^{-1}$ (Fig.~\ref{fig:Library}c) yields the geometric design rule

\begin{equation}
    w_1=g^{-1}\!\left(f\!\left(g(w_2)\right)\right),
    \label{eqs}
\end{equation}
where $w_1$ and $w_2$ are the side widths of the upper and lower nanopillars, respectively. Once $w_2$ is chosen to span the required $0$--$2\pi$ transmission phase, equation (\ref{eqs}) prescribes the value of $w_1$ that minimizes reflection for each meta-atom.

Fig.~\ref{fig:Library}d,e compare the resulting bilayer library ($d_2=1300$~nm, $d_1=290$~nm) with a conventional single-layer TiO$_2$ reference (pillar height 1300~nm). Both libraries provide full phase coverage, but the bilayer library maintains markedly higher transmittance across the entire set of meta-atoms. In particular, the minimum transmittance increases from 86.7\% in the single-layer library to 97.6\% in the bilayer library. This library-level improvement suppresses reflection without sacrificing phase coverage and should translate into higher device-level efficiency.

This construction relies on the effective-medium approximation, which holds while the lattice response is dominated by the fundamental Bloch mode. We verified that all meta-atoms in the library operate in this single-mode regime by comparing transfer-matrix and RCWA transmittances (Supplementary Section~S3).

The framework is not restricted to identical materials in the two layers; an example hybrid TiO$_2$/a-Si design at $\lambda=1550$~nm is presented in Supplementary Section~S4.

\section*{Device-level validation of bilayer metalenses}\label{subsecfullwave}

We next evaluate whether the antireflective advantage obtained at the meta-atom-library level is retained in a complete optical device. We compare metalenses designed from the bilayer library with reference metalenses designed from the single-layer library. Each lattice site of pitch $U$ is assigned the geometry whose transmission phase most closely matches the local target. For a monochromatic metalens focusing into the substrate, the target phase profile is
\begin{equation}
    \varphi(x,y)=n_{\mathrm{sub}}\frac{2\pi}{\lambda_d}
    \left(f-\sqrt{x^2+y^2+f^2}\right),
    \label{eqlensphase}
\end{equation}
where $f$ is the focal length, $\lambda_d$ is the design wavelength, $n_{\mathrm{sub}}$ is the substrate refractive index, and the optical axis is along $z$.

\begin{figure}[h]
\centering
\includegraphics[width=1.00\linewidth]{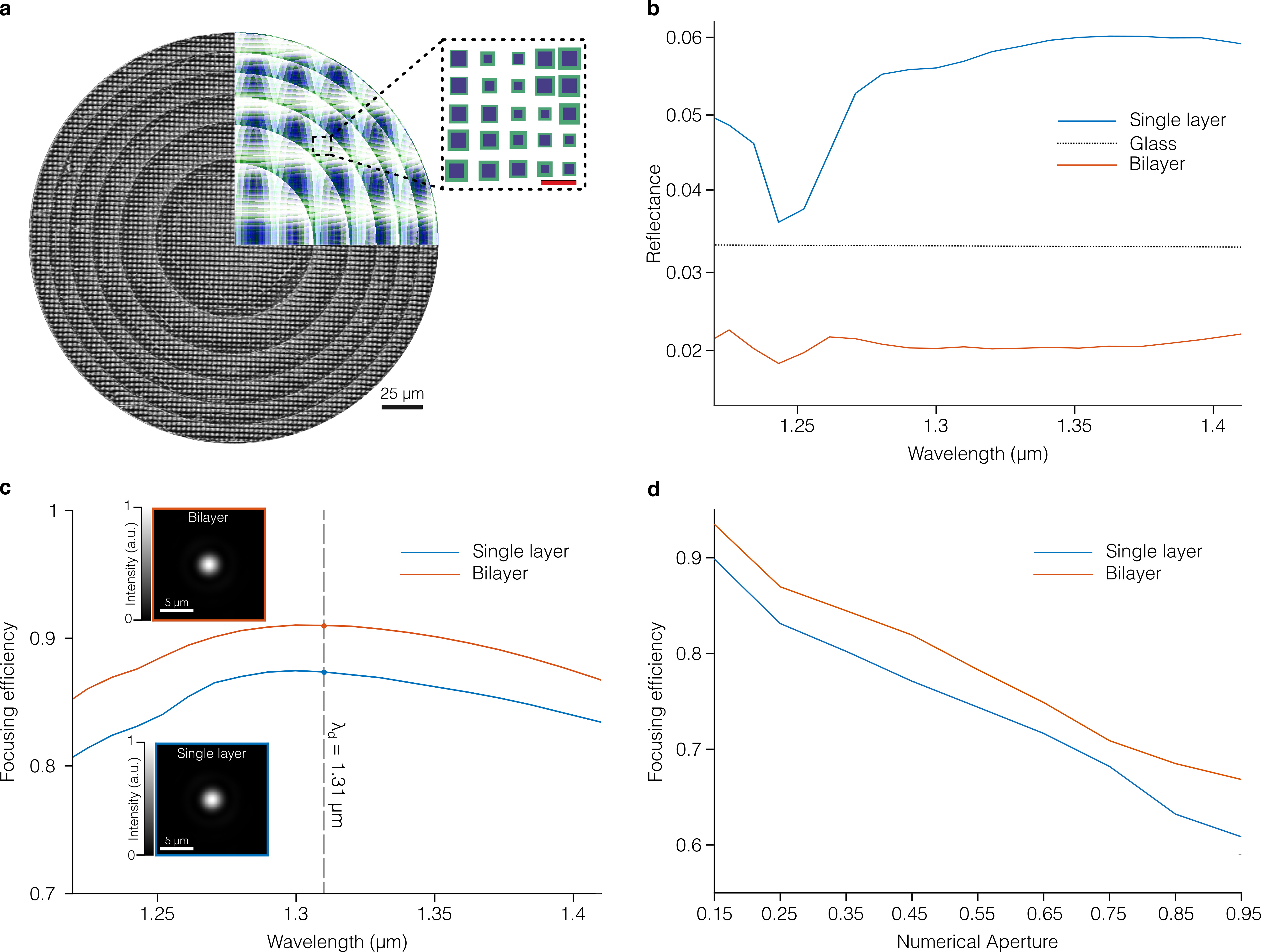}
\caption{\textbf{Full-wave validation of TiO$_2$/TiO$_2$ antireflective bilayer metalenses.}
\textbf{a,} SEM picture of a fabricated bilayer metalens overlaid with a bird's-eye rendering of one quarter of a representative device designed for $\lambda_d=1310$~nm, focal length 1 mm, diameter 250~$\mu$m and numerical aperture 0.18, focusing into the glass substrate. Insets show the lower TiO$_2$ pillars, adjacent to the substrate, in green and the upper TiO$_2$ pillars, adjacent to air, in blue (scale bar is 700 nm).
\textbf{b,} Simulated reflectance spectra of the bilayer metalens, the single-layer reference metalens and a bare glass substrate. At the design wavelength, the bilayer architecture reduces the reflectance below both the single-layer device and bare glass.
\textbf{c,} Simulated focusing-efficiency spectra of the bilayer and single-layer metalenses. The focusing efficiency is obtained by integrating the focal-plane PSF over a circular region centred on the focus whose radius is set by the third zero of the diffraction-limited Airy pattern, $\rho_3=u_3\lambda/(2\pi\mathrm{NA})$, with $u_3=10.1735$. The result is normalized by the power transmitted through the same glass substrate without the patterned metalens. Insets show the corresponding PSFs at the design wavelength.
\textbf{d,} Simulated focusing efficiency of bilayer and single-layer metalenses designed over a range of numerical apertures. The bilayer architecture yields higher simulated device-level efficiency across the full numerical-aperture range.}
\label{fig:FWSim}
\end{figure}

Figure~\ref{fig:FWSim}a shows a representative bilayer metalens designed for $\lambda_d=1310$~nm, with focal length $f=1~$mm, diameter $D=250~\mu$m and NA 0.18. The lower TiO$_2$ pillars are shown in green, whereas the upper TiO$_2$ pillars in blue. A single-layer TiO$_2$ metalens with identical aperture and focal length serves as the reference.

We simulated both devices using full-aperture finite-difference time-domain calculations in Tidy3D~\cite{hughes2021perspective}, with the actual pillar geometries over the complete 250~$\mu$m aperture rather than a unit-cell approximation. The simulated reflectance spectra are shown in Fig.~\ref{fig:FWSim}b. At the design wavelength, the reflectance decreases from 5.6\% for the single-layer device to 2\% for the bilayer device — less than the 3.3\% Fresnel reflectance of a bare glass substrate. The bilayer architecture suppresses the reflected contributions from all three interfaces of the effective stack, whereas a top coating addresses only the air--metasurface interface.

Let $P_{\mathrm{in}}$ denote the incident power and $P_{\mathrm{focus}}$ the power integrated over a circular region of diameter $\sim 3.24\,\lambda/\mathrm{NA}$ centred on the focus, corresponding to the third zero of the diffraction-limited Airy pattern (see
Methods)~\cite{engelberg2020near}. With $R_{\mathrm{glass}} \approx 3.3\%$ the Fresnel reflectance of the bare substrate, the substrate-normalized focusing efficiency is
\begin{equation}
    \eta_{\mathrm{focus}} = \frac{P_{\mathrm{focus}}}{P_{\mathrm{ref}}} 
    = \frac{P_{\mathrm{focus}}}{(1 - R_{\mathrm{glass}})\,P_{\mathrm{in}}},
    \label{eq:eta_focus}
\end{equation}
which matches the experimental normalization used below. The absolute focusing efficiency, which includes reflection losses at the metasurface itself, is
\begin{equation}
    \eta_{\mathrm{abs}} = \eta_{\mathrm{focus}}\,(1 - R_{\mathrm{glass}}),
    \label{eq:eta_abs}
\end{equation}
and is directly obtainable from the measured $\eta_{\mathrm{focus}}$.

Figure~\ref{fig:FWSim}c shows the focusing-efficiency spectra. At the design wavelength, the bilayer device reaches 91.1\%, compared with 87.4\% for the single-layer reference. The corresponding PSFs (insets of Fig.~\ref{fig:FWSim}c, Strehl ratios $\sim 0.99$) are nearly indistinguishable, so the two devices provide comparable wavefront fidelity while the bilayer architecture delivers more power to the focus.

These results confirm the central premise of the design: antireflection and diffraction-limited wavefront control are achieved simultaneously within the same meta-atom architecture. A loss-channel decomposition (Supplementary Section~S5) further confirms that the efficiency enhancement is dominated by increased throughput rather than improved wavefront formation.

To test whether this advantage persists for wavefronts with steeper phase gradients, we simulated metalenses with NAs from 0.15 to 0.95. As shown in Fig.~\ref{fig:FWSim}d, the bilayer architecture consistently outperforms the single-layer design across the full range. The same efficiency advantage is observed for a hybrid TiO$_2$/a-Si bilayer metalens at NA~$= 0.36$ operating at $\lambda_d = 1550$  nm, confirming that the strategy is not restricted to the TiO$_2$/TiO$_2$ platform (Supplementary Section~S6).

In Supplementary Section~S7, we further examine the tolerance of a TiO$_2$/TiO$_2$ bilayer metalens to interlayer misalignment; reflectance remains below that of bulk glass over a wide range of lateral offsets.

\section*{Bilayer fabrication platform}\label{subsecFab}

A key requirement of the antireflective bilayer strategy is independent control over the lateral geometry of the two metasurface layers. The bilayer design requires independent in-plane patterning of the two layers with accurate vertical registration between them.

\begin{figure}[h]
\centering
\includegraphics[width=1.00\linewidth]{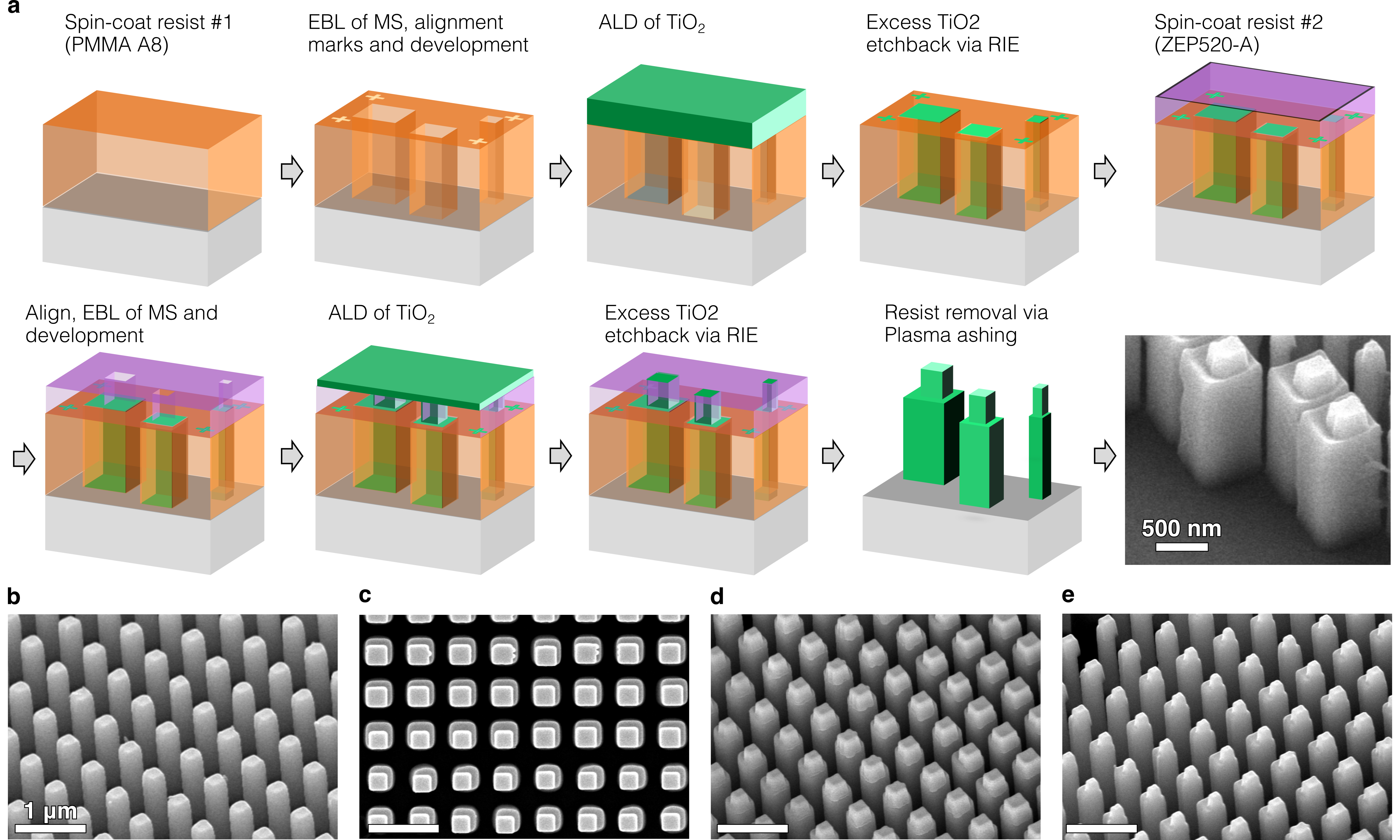}
\caption{\textbf{Bilayer fabrication platform for independently controlled TiO$_2$ metasurfaces.}
\textbf{a,} Process flow used to fabricate the bilayer device. A first resist layer is patterned, filled with TiO$_2$ by atomic layer deposition, and planarized by reactive-ion etch-back to define the lower nanopillar layer. A second resist layer is then aligned, patterned independently, filled with TiO$_2$, and planarized in the same way, followed by resist removal to yield free-standing bilayer pillars. A representative SEM of the resulting structures is shown at right.
\textbf{b,} SEM image of the fabricated single-layer device.
\textbf{c,} Top-view image of the bilayer. One can appreciate the vertical alignment of the upper and lower TiO$_2$ layers across the device.
\textbf{d,e,} Higher-magnification tilted SEM of the bilayer nanopillar arrays, showing independent definition and uniformity. (Scale bar is 1 $\mu$m for b-e images)}
\label{fig:Fab}
\end{figure}

We fabricated the single-layer and bilayer metalenses using single- and double-damascene processes, respectively~\cite{devlin2016broadband, chen2018broadband} (Fig.~\ref{fig:Fab}a; see Methods). The bottom-up damascene approach intrinsically preserves vertical sidewalls and independent lateral definition of each layer, avoiding the taper and profile nonidealities that can arise in deep etching processes. The first TiO$_2$ layer is 1300~nm tall, with aspect ratios of up to $\sim 13$, and the combined bilayer thickness reaches nearly 1600~nm — among the tallest high-aspect-ratio TiO$_2$ meta-atoms reported for metalens implementations. After deposition and etch-back of the second layer, the resist stack is removed by oxygen plasma ashing, yielding the free-standing bilayer pillars shown in Fig.~\ref{fig:Fab}b--e. The micrographs confirm the quality of the process: the lower- and upper-layer pillars are clearly resolved with vertical sidewalls (Fig.~\ref{fig:Fab}d,e), the arrays are uniform across the device, and the top-view image (Fig.~\ref{fig:Fab}c) shows that the two layers remain centered on one another, confirming the accuracy of the interlayer alignment.

The single-layer and bilayer metalenses were fabricated on the same substrate to ensure identical first-layer geometry and post-processing conditions; the only difference is that the single-layer device was not exposed during the second lithography step. This isolates the effect of the bilayer architecture from run-to-run fabrication variation.

\section*{Experimental characterization of antireflective metalenses}\label{subsecExp}

\begin{figure}[h]
\centering
\includegraphics[width=1.00\linewidth]{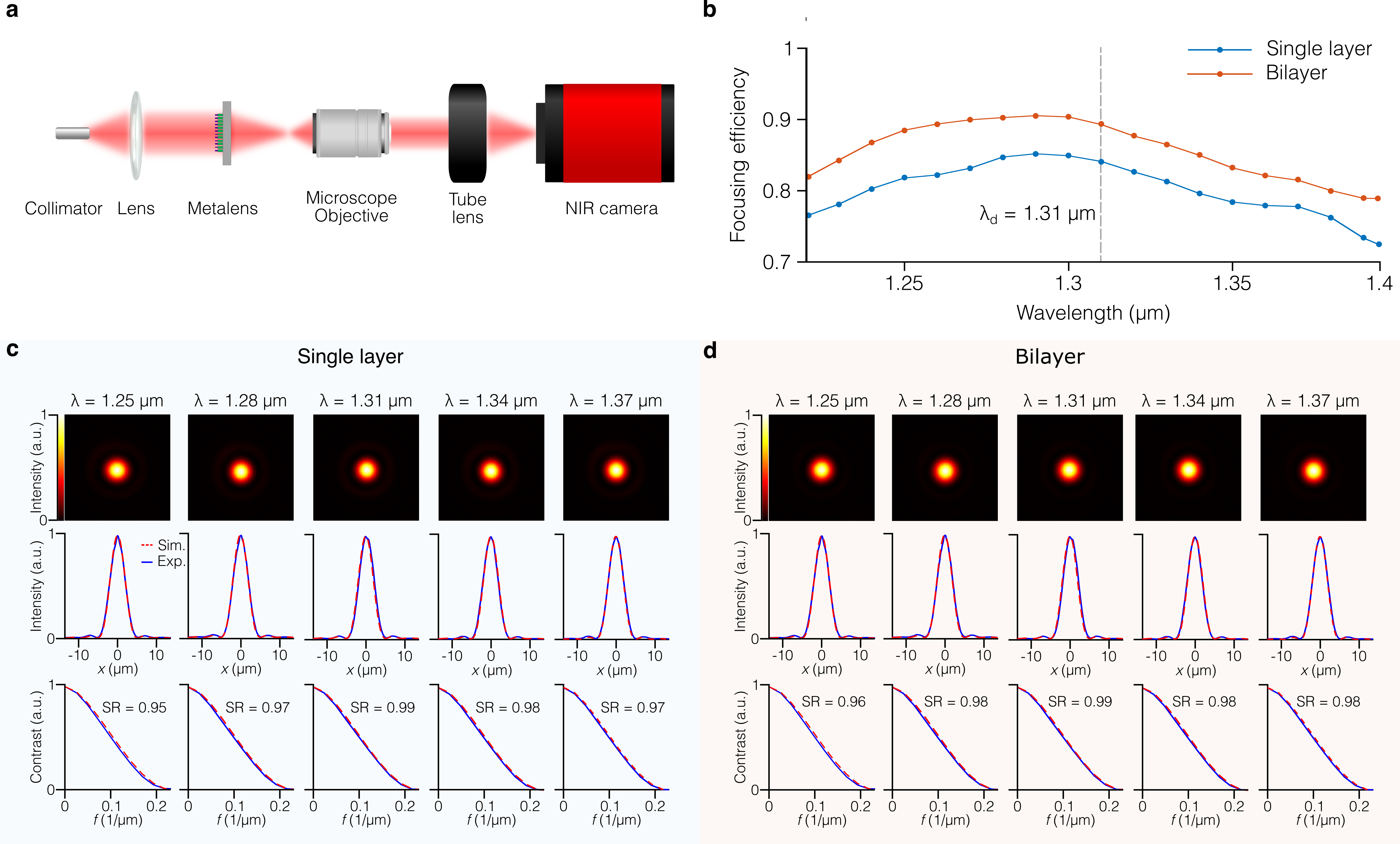}
\caption{\textbf{Experimental validation of single-layer and bilayer TiO$_2$ metalenses.}
\textbf{a,} Schematic of the optical characterization setup. Light from a tunable laser is collimated and focused onto the metalens aperture. The transmitted focal field is collected with a $50\times$ objective, relayed by a tube lens and recorded on a near-infrared camera.
\textbf{b,} Measured focusing efficiency as a function of wavelength for the single-layer and bilayer metalenses. The focusing efficiency is obtained by integrating the measured focal-plane intensity over a circular region centred on the focus whose radius is set by the third zero of the diffraction-limited Airy pattern, and normalizing by the power transmitted through the same glass substrate without the patterned metalens. The bilayer device maintains higher efficiency across the measured spectral range and reaches 90\% near the design wavelength, compared with 86\% for the single-layer control.
\textbf{c,d,} Measured focusing performance of the single-layer \textbf{(c)} and bilayer \textbf{(d)} metalenses at representative wavelengths. For each wavelength, the top row shows the measured focal-plane intensity distribution (field of view $20\,\mu$m $\times$ $20\,\mu$m, matching the horizontal scale of the plots below), the middle row shows the corresponding PSF line cut compared with the simulated reference, and the bottom row shows the modulation 
transfer function (MTF). The SR values indicate the Strehl ratios extracted from the measured focal spots.}
\label{fig:Exp}
\end{figure}

We fabricated and characterized two TiO$_2$-on-glass metalenses, a  conventional single-layer device and an antireflective bilayer device, both with the same target parameters as the simulated devices. The measurement setup is shown in Fig.~\ref{fig:Exp}a (see Methods for details).

The measured focusing efficiencies are shown in Fig.~\ref{fig:Exp}b. The bilayer metalens outperforms the single-layer device over a bandwidth of approximately 200~nm, reaching 90\% near the design wavelength compared with 86\% for the single-layer control. Because the two devices share the same aperture, focal length and numerical aperture, this efficiency difference reflects an increase in the fraction of incident power delivered to the focus.

The measured focal profiles at representative wavelengths are shown in Fig.~\ref{fig:Exp}c,d. For each device and wavelength, we compare the focal-plane intensity distribution, a line cut through the point-spread function (PSF), and the corresponding modulation transfer function (MTF). Both devices remain diffraction-limited across the measured spectral range. At the design wavelength, the Strehl ratio reaches approximately 0.99 and remains above 0.95 across the measured range, confirming that the bilayer architecture preserves the wavefront fidelity of the single-layer reference.

The efficiency gain is therefore not a consequence of improved phase control but of reduced reflection. The measured $\sim$4 percentage-point efficiency increase agrees with the simulated 3.6~percentage points reflectance reduction (5.6\% to 2\%), providing quantitative experimental evidence — together with the indistinguishable focal profiles — that the bilayer architecture improves efficiency by suppressing parasitic reflection. A direct reflectance measurement was not pursued because the metasurface scatters reflected light over a wide angular range, which a finite-aperture detector cannot capture without systematic collection errors.

\section*{Conclusions}\label{secConclusion}

Antireflection coatings have traditionally been designed as optical layers added to an interface after the function of the underlying element has been defined. This separation between impedance matching and wavefront control is poorly suited to metasurfaces, where each phase state is implemented by a different meta-atom geometry and therefore presents a different optical impedance. Here we have shown that bilayer metasurfaces provide a route to remove this separation. By treating each vertically integrated meta-atom as both a phase-shifting element and an effective antireflective stack, reflection can be addressed at the level of the phase library itself rather than as a post-fabrication correction.

Using this principle, we designed antireflective TiO$_2$/TiO$_2$ bilayer libraries with full $0$--$2\pi$ phase coverage and implemented them in near-infrared metalenses that suppress reflection below bare glass and increase focusing efficiency relative to single-layer controls while preserving diffraction-limited imaging. While modest in isolation, the roughly 4 percentage-point efficiency gain compounds multiplicatively in cascaded metasurface systems for imaging, holography or optical computing~\cite{dorrah2025compound}, where every percentage point of throughput at each surface translates directly into system-level performance.

The framework introduced here assumes normal incidence and a fixed design wavelength, but both constraints can in principle be relaxed within the same architecture. The thin-film model extends naturally to oblique incidence by replacing the Fresnel coefficients with their angle-dependent counterparts, opening a route to wide field-of-view designs. Independently, the upper-layer geometry provides an additional dispersion handle that may be co-optimized for antireflection and group-delay control in achromatic metalenses, a trade-off that has limited single-layer designs~\cite{li2024heterogeneous}. More generally, the framework is not limited to scalar focusing or identical-material bilayers: the same principle extends naturally to hybrid material systems, polarization-dependent  responses, and multifunctional devices. By embedding antireflection directly into the meta-atom architecture, bilayer metasurfaces offer a path towards flat-optical systems in which high throughput, chromatic control \cite{presutti2020focusing} and complex wavefront response are engineered together from the outset.

\section*{Methods}\label{secMethods}

\subsection*{Fabrication of single-layer and bilayer TiO$_2$ metalenses}

Single-layer and bilayer metalenses were fabricated on 500-$\mu$m-thick fused-silica substrates using single- and double-damascene processes, respectively~\cite{devlin2016broadband, chen2018broadband}. The first TiO$_2$ layer was defined by electron-beam lithography in 950 PMMA A8, filled by atomic layer deposition at 90~$^\circ$C, and planarized by reactive-ion etch-back. For bilayer devices, this sequence was repeated using ZEP520A as the second resist, developed in cold \textit{o}-xylene at 5~$^\circ$C to avoid degradation of the underlying TiO$_2$ structures~\cite{dorrah2025free}, followed by oxygen plasma ashing to yield free-standing bilayer pillars. 

Both device types were fabricated on the same substrate, with the single-layer control undergoing the full second-layer processing sequence but without pattern exposure, ensuring identical first-layer fabrication and post-processing conditions for both devices.

\subsection*{Optical characterization and efficiency measurements}\label{subsecMeas}

The optical characterization setup is shown in Fig.~\ref{fig:Exp}a. Light from a tunable source (SuperK laser, NKT Photonics; LLTF filter, Photon etc.) was collimated and focused onto the metalens aperture. The transmitted focal field was collected with a Mitutoyo M Plan Apo NIR $50\times$ objective (NA~0.42), relayed by a tube lens, and recorded on a Raptor OWL 640 near-infrared camera.

Focusing efficiencies were measured using a substrate-normalized reference, following Ref.~\cite{engelberg2020near}. The substrate-normalized reference power was obtained by translating the sample to an unpatterned region of the same substrate and recording the transmitted beam profile under identical illumination conditions. The focal power was integrated over a circular aperture of diameter $D_3 \simeq 3.24\,\lambda/\mathrm{NA}$, corresponding to the third zero of the Airy pattern, and normalized by the reference intensity integrated over a 250~$\mu$m circular mask matching the device aperture. Reported values are averages over 100 camera frames. Strehl ratios were extracted from the peak PSF intensity after normalization to the diffraction-limited Airy profile at the same numerical aperture.

\backmatter

\bmhead{Supplementary information}

The data supporting the findings of this study are available in the
Article and its Supplementary Information. 

\bmhead{Acknowledgements}

The authors from Harvard University acknowledge financial support from Corning Incorporated. This work was performed in part at the Harvard University Center for Nanoscale Systems (CNS), a member of the National Nanotechnology Coordinated Infrastructure Network (NNCI), which is supported by the National Science Foundation under NSF award no. ECCS-2025158. Part of the simulations presented in this paper were performed using Tidy3D. Part of the computations in this paper were run on the FASRC Cannon cluster supported by the FAS Division of Science Research Computing Group at Harvard University. D.C., A.P., and F.C. acknowledge funding support from the Air Force Office of Scientific Research under Award FA9550-21-1-0312. We thank M. Yessenov for the fruitful discussions and feedback. The authors also thank J.S. Park for helpful discussions on fabrication. 

\bmhead{Author contributions}
J.O., D.C., A.P. and P.D. conceived the idea. J.O. performed the thin-film model simulations, meta-atom library design and metalens simulations. D.C. performed metalens simulations, fabricated the devices and carried out the optical measurements. A.P. performed metalens simulations and carried out the optical measurements. L.P. carried out the optical measurements. J.R.R. and A.H.D. participated in discussions. P.D. and F.C. supervised the project, provided resources, and participated in discussions. J.O., D.C. and A.P. wrote the manuscript with input from all authors.

\noindent

\bibliography{sn-bibliography}

\end{document}


\begin{center}
{\Large \textbf{Supplementary Information}}\\[0.6em]
{\large \textbf{Antireflection by design in bilayer metasurfaces}}\\[0.8em]
Jaewon Oh$^{1,2,\dagger,*}$, Davide Cassara$^{1,\dagger}$, Alfonso Palmieri$^{1,\dagger}$,
Lorenzo Piatti$^{1}$, Janderson Rocha Rodrigues$^{2}$,
Ahmed H. Dorrah$^{1,3}$, Paulo Dainese$^{2}$ and Federico Capasso$^{1,*}$\\[0.8em]
$^{1}$Harvard John A. Paulson School of Engineering and Applied Sciences, Harvard University, Cambridge, Massachusetts 02138, USA\\
$^{2}$Corning Research and Development Corporation, Painted Post, New York 14870, USA\\
$^{3}$Applied Physics and Science Education, Eindhoven University of Technology, Eindhoven, The Netherlands\\
$^\dagger$These authors contributed equally to this work\\
$^*$Correspondence: \texttt{ohj7@corning.com, capasso@seas.harvard.edu}
\end{center}

\tableofcontents
\clearpage

\section{Vector-method and transfer-matrix descriptions of the effective bilayer stack}
\label{sec:vector_tmm}

\begin{figure}[h]
\centering
\suppfigure{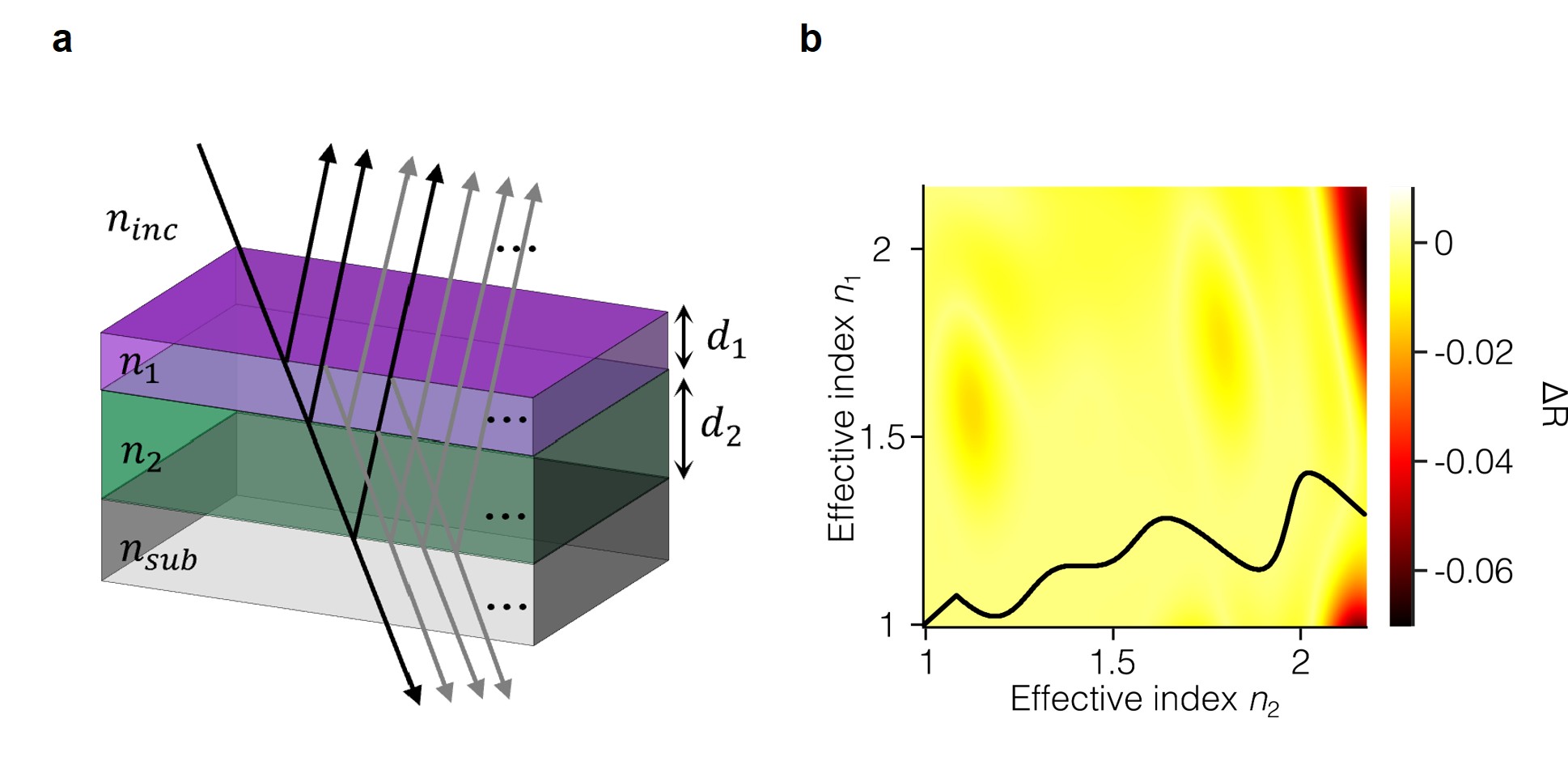}{0.90\linewidth}
\caption{\textbf{Comparison between the vector method and the transfer-matrix method.}
\textbf{a,} Dominant reflected contributions from the three interfaces are shown explicitly. Gray rays
indicate representative higher-order paths generated by multiple internal reflections.
\textbf{b,} Reflectance difference
$\Delta R=R_{\mathrm{TMM}}-R_{\mathrm{vec}}$ between the transfer-matrix and vector-method
calculations, plotted on a linear reflectance scale as a function of the effective indices
$n_2$ and $n_1$ for $d_1=\SI{290}{nm}$, $d_2=\SI{1300}{nm}$ and
$\lambda=\SI{1310}{nm}$. The black curve indicates the antireflective design locus
$f(n_2)=n_1$ used to construct the TiO$_2$/TiO$_2$ bilayer library.}
\label{fig:S1}
\end{figure}

The thin-film model used in the main text represents each bilayer meta-atom as an effective
air/film~1/film~2/substrate stack. In this model, the two patterned nanopillar layers are
replaced by homogeneous films with effective refractive indices $n_1$ and $n_2$ and
thicknesses $d_1$ and $d_2$, respectively. The incident medium has refractive index
$n_{\mathrm{inc}}$, and the substrate has refractive index $n_{\mathrm{sub}}$.

The vector method used in the main text approximates the reflected field by retaining the
first reflected contribution from each of the three effective interfaces, as shown in
\cref{fig:S1}a. This approximation provides a simple geometric interpretation of antireflection. Exact cancellation occurs when the three reflected phasors form a closed
triangle in the complex plane; low reflectance corresponds to a nearly closed triangle \cite{macleod}.
The vector method therefore gives a simple design rule for locating antireflective regions
in the $(n_1,n_2)$ index space. However, it neglects higher-order reflected fields produced
by multiple internal reflections within the two films. To assess the accuracy of this
approximation over the design space used in the bilayer library, we compare it with the
transfer-matrix method, which accounts for the complete sequence of forward- and
backward-propagating waves in the stack.\\
At normal incidence, the tangential electric and magnetic fields at the two sides of layer
$m$ are related by the characteristic matrix
\begin{equation}
\begin{bmatrix}
E\\
H
\end{bmatrix}_{m,\mathrm{in}}
=
\mathbf{M}_m
\begin{bmatrix}
E\\
H
\end{bmatrix}_{m,\mathrm{out}},
\qquad
\mathbf{M}_m=
\begin{bmatrix}
\cos\delta_m & i y_m^{-1}\sin\delta_m\\
i y_m \sin\delta_m & \cos\delta_m
\end{bmatrix},
\label{eq:S_characteristic_matrix}
\end{equation}
where
\begin{equation}
\delta_m = \frac{2\pi}{\lambda}n_m d_m
\label{eq:S_phase_thickness}
\end{equation}
is the phase thickness of the layer. Here $n_m$ and $d_m$ are the effective refractive
index and thickness of layer $m$, respectively. The quantity
$y_m=n_m\mathcal{Y}_0$ is the optical admittance of the layer for non-magnetic media
at normal incidence, with $\mathcal{Y}_0$ the free-space admittance.\\
For the two-layer stack, the fields at the incident side of the structure are related to
those at the substrate side by
\begin{equation}
\begin{bmatrix}
E\\
H
\end{bmatrix}_{a}
=
\mathbf{M}_1\mathbf{M}_2
\begin{bmatrix}
E\\
H
\end{bmatrix}_{s},
\label{eq:S_total_matrix}
\end{equation}
where $a$ denotes the incident side and $s$ denotes the semi-infinite substrate. Since the
substrate supports only a forward-propagating transmitted wave,
\begin{equation}
\begin{bmatrix}
E\\
H
\end{bmatrix}_{s}
=
\begin{bmatrix}
1\\
y_{\mathrm{sub}}
\end{bmatrix}E_s .
\label{eq:S_substrate_boundary}
\end{equation}
Defining \(B\) and \(C\) as the electric and magnetic fields at the incident side
scaled by the transmitted electric-field amplitude in the substrate,
\begin{equation}
B = \frac{E_a}{E_s}, \qquad C = \frac{H_a}{E_s},
\label{eq:S_BC_definition}
\end{equation}
and substituting equation~\eqref{eq:S_substrate_boundary} into
equation~\eqref{eq:S_total_matrix}, we obtain
\begin{equation}
\mathbf{M}_1\mathbf{M}_2
\begin{bmatrix}
1\\
y_{\mathrm{sub}}
\end{bmatrix}
=
\begin{bmatrix}
B\\
C
\end{bmatrix}.
\label{eq:S_BC_matrix}
\end{equation}
The effective input admittance of the complete stack is then
\begin{equation}
Y_{\mathrm{in}}=\frac{C}{B}.
\label{eq:S_input_admittance}
\end{equation}
The amplitude reflection coefficient and reflectance obtained from the transfer-matrix
method are therefore
\begin{equation}
r_{\mathrm{TMM}}=
\frac{y_{\mathrm{inc}}-Y_{\mathrm{in}}}{y_{\mathrm{inc}}+Y_{\mathrm{in}}},
\qquad
R_{\mathrm{TMM}}=\left|r_{\mathrm{TMM}}\right|^2 ,
\label{eq:S_tmm_reflectance}
\end{equation}
where $y_{\mathrm{inc}}=n_{\mathrm{inc}}\mathcal{Y}_0$ is the optical admittance of the
incident medium.\\
For comparison, the vector-method reflectance used in the main text is
\begin{equation}
R_{\mathrm{vec}}(n_1,n_2,d_1,d_2)=
\left|
r_0+r_1 e^{2 i\delta_1}+r_2 e^{2 i(\delta_1+\delta_2)}
\right|^2,
\label{eq:S_vector_reflectance}
\end{equation}
with
\begin{equation}
r_0=\frac{n_{\mathrm{inc}}-n_1}{n_{\mathrm{inc}}+n_1},\qquad
r_1=\frac{n_1-n_2}{n_1+n_2},\qquad
r_2=\frac{n_2-n_{\mathrm{sub}}}{n_2+n_{\mathrm{sub}}}.
\label{eq:S_fresnel_coefficients}
\end{equation}
Here $r_0$, $r_1$ and $r_2$ are the Fresnel reflection coefficients associated with the
incident-medium/film~1, film~1/film~2 and film~2/substrate interfaces, respectively.
The exponential factors account for the phase accumulated by the reflected fields before
they recombine at the incident side of the stack.\\
\Cref{fig:S1}b compares $R_{\mathrm{vec}}$ and $R_{\mathrm{TMM}}$ for the effective
TiO$_2$/TiO$_2$ bilayer stack used in the main-text design, with
$d_1=\SI{290}{nm}$, $d_2=\SI{1300}{nm}$ and $\lambda=\SI{1310}{nm}$. We plot the difference
\begin{equation}
\Delta R =
 R_{\mathrm{TMM}}-R_{\mathrm{vec}},
\label{eq:S_reflectance_error}
\end{equation}
where both reflectances are evaluated on a linear scale. The black curve indicates the
antireflective design locus $f(n_2)=n_1$ used to construct the bilayer library. Along this
locus, the mean difference between the vector-method and transfer-matrix reflectances is
approximately $5\times 10^{-4}$. This small discrepancy shows that, although the vector
method is approximate, it accurately identifies the low-reflection design path relevant
to the meta-atom library while preserving the simple phasor picture used to guide the
bilayer design.

\clearpage

\section{Thickness selection and vector-triangle verification}
\label{sec:triangle}

\begin{figure}[h]
\centering
\suppfigure{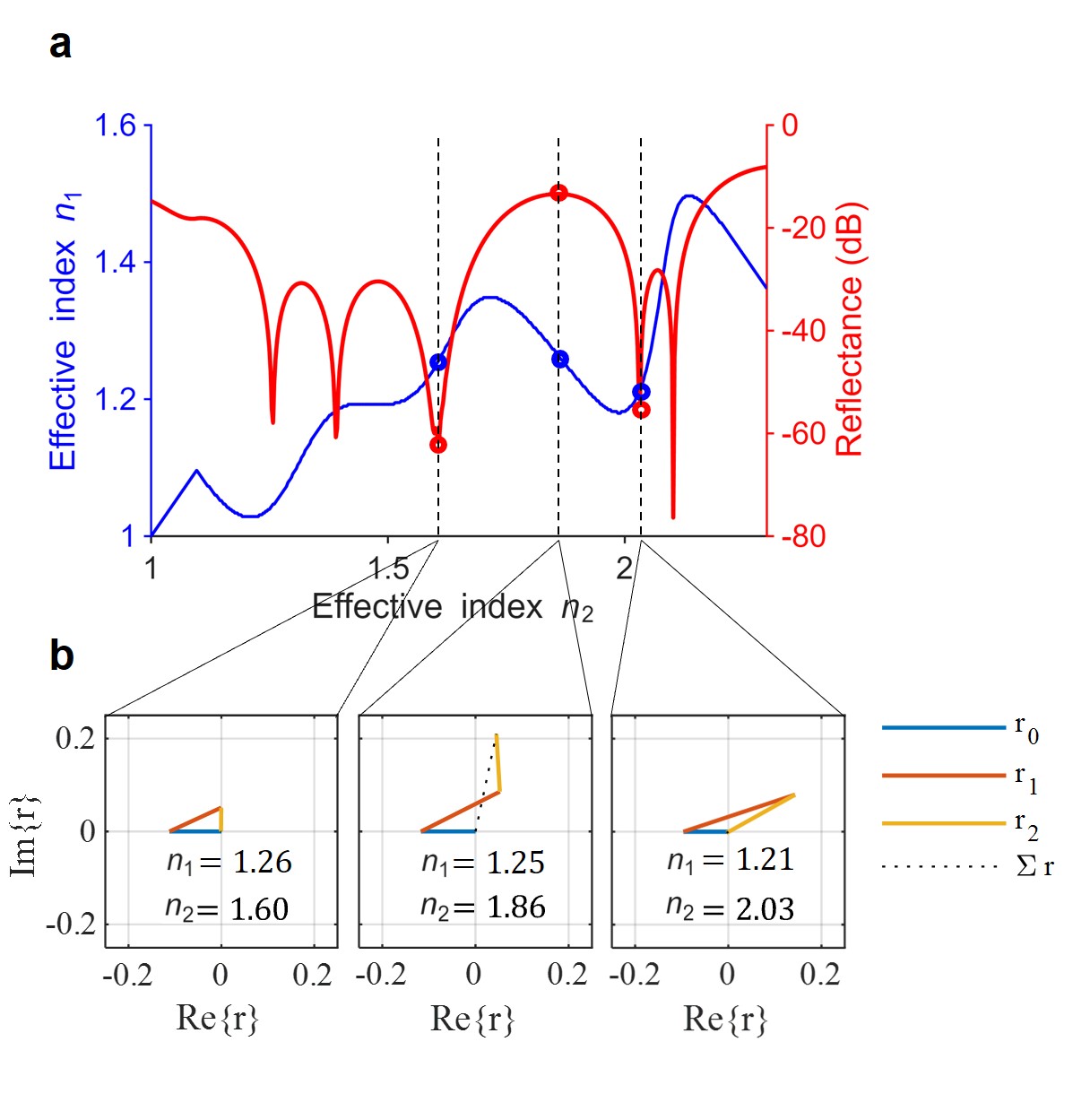}{0.6\linewidth}
\caption{\textbf{Antireflection mapping and vector-triangle verification. } \textbf{a,} Mapping function $f(n_2) = n_1$ (blue, left axis) and minimized reflectance $R_\mathrm{min}$ (orange, right axis) as a function of the lower-layer effective index $n_2$, evaluated at the selected thickness pair $(d_1, d_2) = (290~\mathrm{nm},\, 1300~\mathrm{nm})$. For each $n_2$, $f(n_2)$ gives the upper-layer effective index that minimizes the reflectance of the bilayer stack. Dashed vertical lines mark three representative index pairs selected for vector-triangle analysis in panel~(b). \textbf{b,} Reflected amplitudes $r_0$, $r_1$ and $r_2$ plotted as vectors in the complex plane for the three index pairs indicated in panel~(a). In each case the vectors form a nearly closed triangle, confirming that the reflectance minimum arises from destructive interference among the three interface contributions. The small residual vector sum $\Sigma r$ is consistent with the non-zero values of $R_\mathrm{min}$ at the corresponding points in panel~(a).}
\label{fig:Snew}
\end{figure}
\noindent
To select the bilayer thicknesses, we define a threshold-based figure of merit (FOM) as the fraction of uniformly sampled lower-layer effective-index values \(n_2\) along the minimized-reflectance locus \(R_\mathrm{min}\) for which the reflectance remains below a threshold \(R_\mathrm{th}\). This metric is more appropriate than an average reflectance because each value of \(n_2\) corresponds to a candidate library element. The objective is therefore to obtain a broad set of antireflective meta-atoms rather than a small number of exceptionally low-reflectance geometries.\\
For \(R_\mathrm{th}=-30~\mathrm{dB}\), the FOM exhibits two local maxima near \((d_1,d_2)=(240~\mathrm{nm},1130~\mathrm{nm})\) and \((290~\mathrm{nm},1300~\mathrm{nm})\). Although these two maxima have comparable peak FOM values, 0.44 and 0.42, respectively, their tolerance to thickness deviations differs substantially. The contiguous region in which the FOM remains within 10\% of its peak value spans \(23~\mathrm{nm}\times67~\mathrm{nm}\) around \((240~\mathrm{nm},1130~\mathrm{nm})\), compared with \(79~\mathrm{nm}\times280~\mathrm{nm}\) around \((290~\mathrm{nm},1300~\mathrm{nm})\). We therefore select the latter thickness pair, whose broader tolerance plateau makes the antireflective response more robust to fabrication-induced layer-height variations.\\
Using this selected thickness pair, \((d_1,d_2)=(290~\mathrm{nm},1300~\mathrm{nm})\), Fig.~\ref{fig:Snew}a shows the mapping function \(f\) (blue, left axis) and the corresponding minimized reflectance \(R_\mathrm{min}\) (orange, right axis) as functions of the lower-layer effective index \(n_2\). For each value of \(n_2\), \(f(n_2)\) gives the upper-layer effective index \(n_1\) that minimizes the reflectance of the effective bilayer stack, as defined in Eqs.~(4) and (5) of the main text. The minimized reflectance \(R_\mathrm{min}\) oscillates along this locus as a function of \(n_2\), reaching values well below \(-30~\mathrm{dB}\) at several points. These low-reflectance regions illustrate the same threshold criterion used in the FOM-based thickness selection described above. Three representative index pairs, marked by dashed vertical lines, are selected for vector-triangle analysis in Fig.~\ref{fig:Snew}b.\\
Fig.~\ref{fig:Snew}b shows the three reflected amplitudes \(r_0\), \(r_1\) and \(r_2\), defined in equation~(1) of the main text, plotted as vectors in the complex plane for each of the three selected index pairs. The left and right panels correspond to index pairs located near reflectance minima in Fig.~\ref{fig:Snew}a. In these cases, the reflected phasors form nearly closed triangles, with the residual vector sum \(\Sigma r\) approaching zero. This confirms that the reflectance minima arise from destructive interference among the reflected contributions from the three effective interfaces.\\
The middle panel, by contrast, corresponds to a local maximum of \(R_\mathrm{min}\) between two minima. Here, the triangle does not close as accurately, and the residual vector sum \(\Sigma r\) is visibly larger, consistent with the elevated reflectance at that point in Fig.~\ref{fig:Snew}a. Importantly, however, the complex-plane scale in all three panels spans only \(\pm 0.2\) along both axes, so even in the least favourable case the absolute magnitude of \(\Sigma r\) remains small. This confirms that the bilayer architecture maintains low reflection across the relevant effective-index range, even at points where the triangle-closure condition is not exactly satisfied.

\clearpage

\section{Bloch-mode validity of the effective-index description}
\label{sec:bloch_modes}

\begin{figure}[h]
\centering
\suppfigure{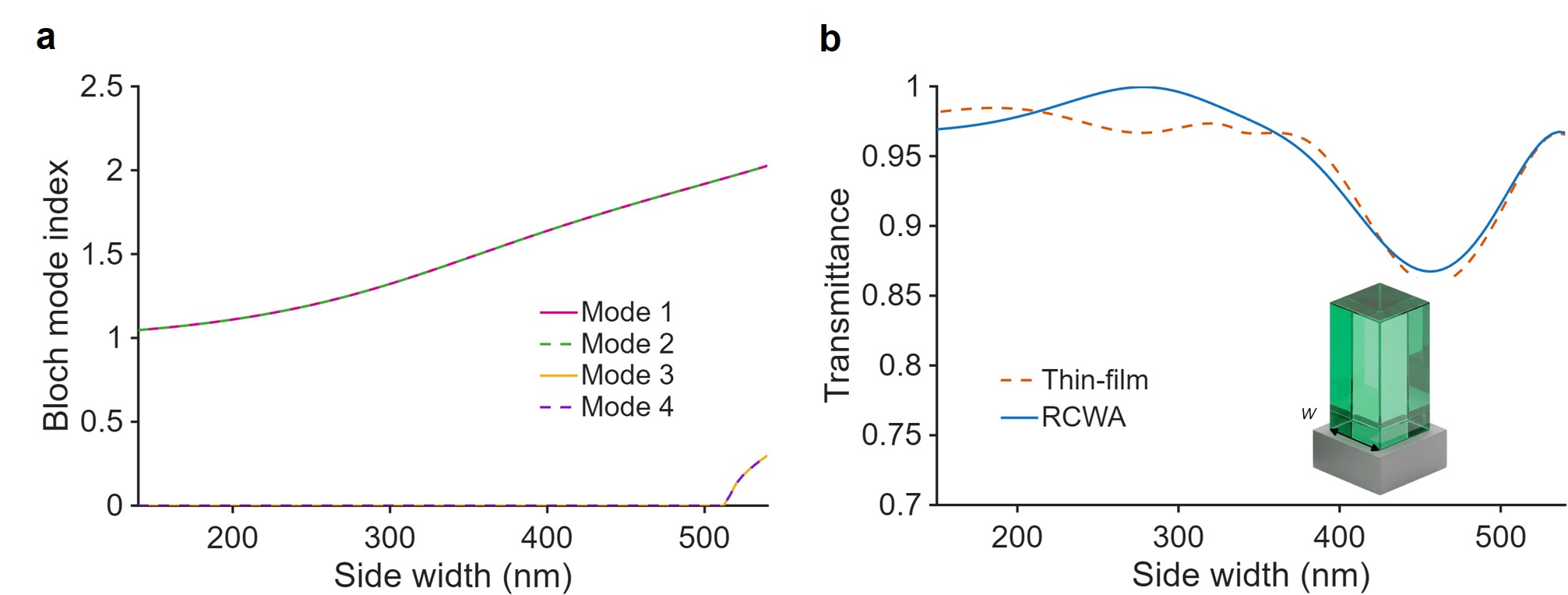}{0.9\linewidth}
\caption{\textbf{Bloch-mode validity of the effective-index description.}
\textbf{a,} Effective indices of the propagating Bloch modes supported by periodic square
TiO$_2$ nanopillars with pitch $U=\SI{600}{nm}$ at $\lambda=\SI{1310}{nm}$,
plotted as a function of side width $w$. The plotted interval,
$w=0.14$--$0.54~\mu\mathrm{m}$, corresponds to the full range used in the meta-atom
library. Higher-order modes become propagating only near the largest side widths,
with an onset at approximately $w=\SI{512}{nm}$.
\textbf{b,} Comparison between the transmittance calculated by full-wave RCWA simulations
of the periodic nanopillar geometry and by an effective thin-film model in which each
meta-atom is represented by an air/effective-film/glass stack with refractive index set by
the fundamental Bloch mode. The mean absolute difference over the full library range is
$\langle |T_{\mathrm{RCWA}}-T_{\mathrm{eff}}| \rangle=0.011$, confirming that the
fundamental-mode effective index remains predictive for the selected geometries.}
\label{fig:S3_bloch}
\end{figure}
\noindent
The antireflective bilayer library construction relies on an effective-index description that maps each nanopillar geometry onto the refractive index of an equivalent thin film. In this approximation, each periodic nanopillar layer is represented by a homogeneous film whose refractive index is set by the propagation constant of the fundamental Bloch mode. For the TiO$_2$/TiO$_2$ platform considered here, the geometry-to-index relation can be written as
\begin{equation}
    g(w)=n_{\mathrm{eff}}(w),
    \label{eq:S3_g}
\end{equation}
where \(w\) is the side width of a square TiO$_2$ nanopillar and \(n_{\mathrm{eff}}\) is the effective index of its fundamental Bloch mode. The thin-film optimization described in the main text and in \cref{sec:vector_tmm} provides the antireflective mapping
\begin{equation}
    n_1=f(n_2),
    \label{eq:S3_f}
\end{equation}
which assigns to each lower-layer effective index \(n_2\) the upper-layer effective index \(n_1\) that minimizes the reflectance of the effective bilayer stack. Combining Eqs.~\eqref{eq:S3_g} and \eqref{eq:S3_f} gives the geometric design rule
\begin{equation}
    w_1=g^{-1}\!\left[f\!\left(g(w_2)\right)\right],
    \label{eq:S3_geometry_rule}
\end{equation}
where \(w_1\) and \(w_2\) are the side widths of the upper and lower nanopillars, respectively. Thus, the lower pillar is selected to provide the desired transmission phase, while the upper pillar is chosen to impedance match the complete bilayer meta-atom.\\
The validity of this construction depends on the assumption that the optical response of the nanopillar lattice is dominated by the fundamental Bloch mode. In general, a periodic lattice of subwavelength pillars can support higher-order Bloch modes, depending on geometry, refractive-index contrast, pitch and wavelength ~\cite{lalanne2006, kikuta1998}. If higher-order modes contributed substantially to the transmitted response, the mapping \(g(w)\) in equation~\eqref{eq:S3_g}, and therefore the design rule in equation~\eqref{eq:S3_geometry_rule}, would no longer provide a reliable description of the full-wave response. The meta-atom dimensions must therefore be restricted to a regime in which the fundamental-mode effective index remains predictive.\\
We examine this condition for the square TiO$_2$ nanopillars over the full side-width interval used in the reference and bilayer libraries. \Cref{fig:S3_bloch}a shows the effective indices of the propagating Bloch modes supported by periodic square TiO$_2$ nanopillars with pitch \(U=\SI{600}{nm}\) at \(\lambda=\SI{1310}{nm}\). The plotted side-width interval, \(w=0.14\)--\(0.54~\mu\mathrm{m}\), corresponds to the full range used in the meta-atom library. This range is set by the unit-cell pitch, fabrication constraints and the requirement of spanning a full \(0\) to \(2\pi\) transmission phase.\\
Over most of this range, the only propagating modes are the two degenerate fundamental modes, as expected from the four-fold symmetry of the square cross-section. Higher-order modes become propagating only near the largest side widths, with their onset occurring at approximately \(w=\SI{512}{nm}\). Their appearance, however, does not by itself imply strong excitation under normal-incidence illumination. The relevant question is whether the fundamental-mode effective index remains predictive of the transmitted response over the full set of geometries used in the library.\\
To test this, we compare two calculations of the single-layer library transmittance. In the first calculation, each TiO$_2$ meta-atom is replaced by an air/effective-film/glass stack. The film thickness is set equal to the TiO$_2$ pillar height, and the film index is taken from the fundamental Bloch mode at the corresponding side width through equation~\eqref{eq:S3_g}. The transmittance of this effective stack is calculated using the transfer-matrix method described in \cref{sec:vector_tmm}. In the second calculation, the transmittance is obtained directly from rigorous coupled-wave analysis (RCWA) of the full periodic nanopillar geometry.\\
As shown in \cref{fig:S3_bloch}b, the effective thin-film model reproduces the RCWA transmittance trend across the full library range. Quantitatively, the mean absolute difference between the two calculations is
\begin{equation}
    \left\langle |T_{\mathrm{RCWA}}-T_{\mathrm{eff}}| \right\rangle = 0.011 .
    \label{eq:S3_T_difference}
\end{equation}
This agreement confirms that coupling to higher-order modes remains weak for the selected geometries, even near the largest side widths where such modes become formally propagating. The TiO$_2$ meta-atoms used in the library can therefore be described, to good approximation, as asymmetric Fabry--Pérot cavities whose cavity index is set by the fundamental Bloch mode. This validates the use of the effective-index mapping \(g(w)\) and of the bilayer geometric design rule in equation~\eqref{eq:S3_geometry_rule} over the full geometry range considered here.

\clearpage

\section{Hybrid TiO$_2$/a-Si antireflective bilayer meta-atoms}
\label{sec:hybrid_library}

The bilayer design demonstrated in the main text uses two TiO$_2$ nanopillar layers.
The underlying principle, however, is not restricted to identical materials. Because the
thin-film model is formulated in terms of effective refractive indices and layer
thicknesses, the two patterned layers can be made from different materials, provided
that their geometries support the required effective-index ranges and remain dominated
by the fundamental Bloch mode.

\begin{figure}[h]
\centering
\suppfigure{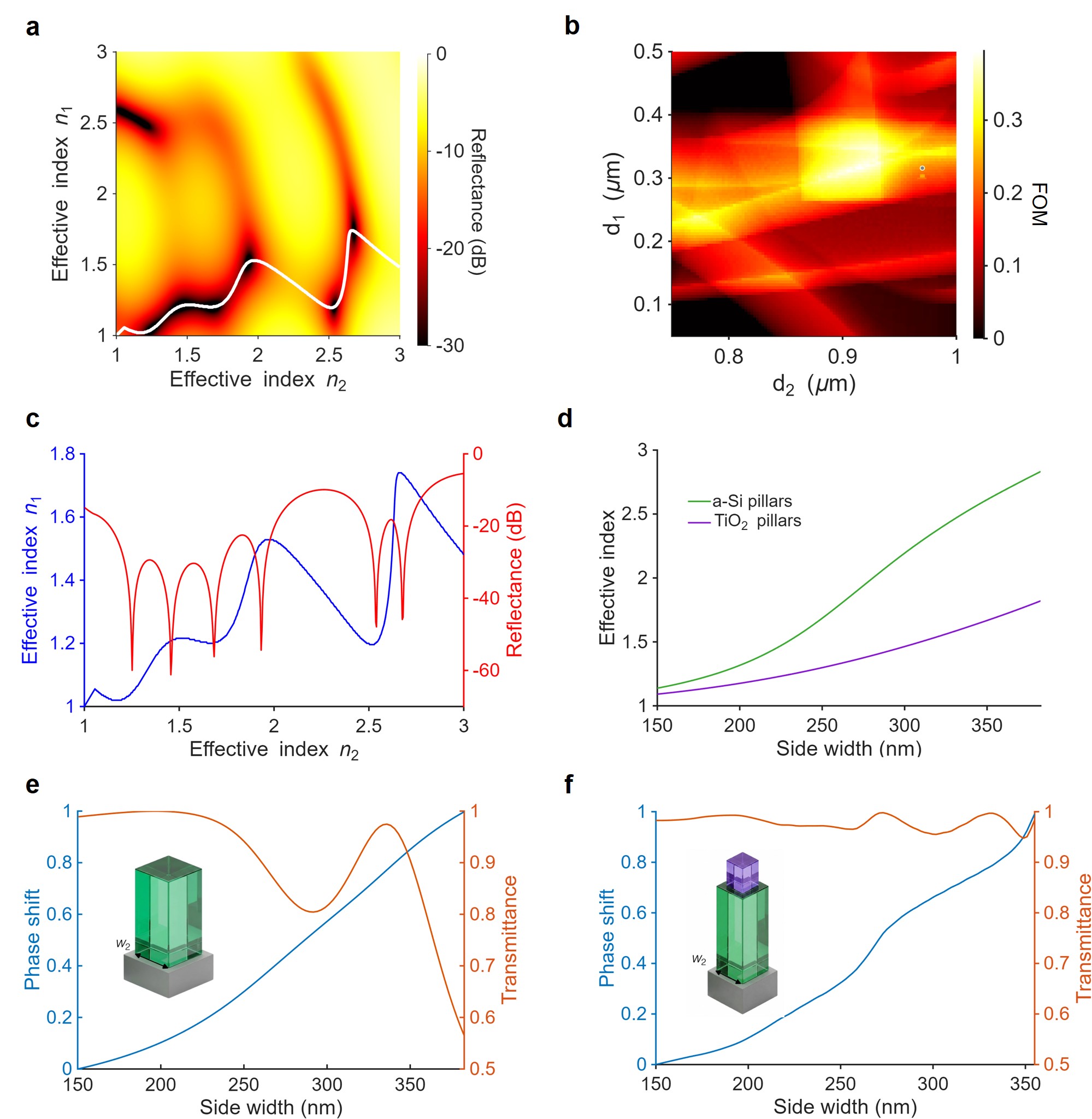}{0.95\linewidth}
\caption{\textbf{Hybrid TiO$_2$/a-Si antireflective bilayer library at
$\lambda=\SI{1550}{nm}$.}
\textbf{a,} Reflectance of an effective air/film~1/film~2/glass stack calculated with the
vector method for the selected thicknesses.
\textbf{b,} Figure of merit evaluated over the thickness parameter space. The selected
thicknesses are $d_1=\SI{320}{nm}$ and $d_2=\SI{910}{nm}$.
\textbf{c,} Mapping function $f(n_2)=n_1$ and corresponding minimized reflectance for
the selected thicknesses.
\textbf{d,} Effective indices of the fundamental Bloch modes supported by periodic square
TiO$_2$ and a-Si nanopillars with pitch $U=\SI{450}{nm}$.
\textbf{e,} Simulated phase shift and transmittance of a conventional single-layer a-Si
meta-atom library. The minimum and mean transmittance are $56.5\%$ and $90.6\%$,
respectively.
\textbf{f,} Simulated phase shift and transmittance of the antireflective TiO$_2$/a-Si
bilayer meta-atom library, with upper TiO$_2$ and lower a-Si heights of
$\SI{320}{nm}$ and $\SI{910}{nm}$, respectively. The minimum and mean transmittance
increase to $94.7\%$ and $97.8\%$, respectively.}
\label{fig:S5}
\end{figure}
\noindent
To illustrate this generality, we designed a hybrid bilayer platform at
$\lambda=\SI{1550}{nm}$ using TiO$_2$ for the upper layer, adjacent to air, and amorphous
silicon (a-Si) for the lower layer, adjacent to the glass substrate. This material ordering
is chosen because a-Si has a high refractive index at telecom wavelengths, enabling
large phase accumulation over a compact lateral footprint, while TiO$_2$ provides an
intermediate-index upper layer that can impedance match the high-index lower layer
to air. The high index of a-Si also allows full $0$ to $2\pi$ phase coverage with a smaller
unit cell, $U=\SI{450}{nm}$, improving the spatial sampling of the target wavefront
relative to the TiO$_2$/TiO$_2$ platform considered in the main text.\\
We first calculated the vector-method reflectance of an effective
air/film~1/film~2/glass stack over the relevant index space and over a range of layer
thicknesses. For each lower-layer effective index $n_2$, we selected the upper-layer
effective index $n_1=f(n_2)$ that minimizes the reflectance. The corresponding
minimized-reflectance curve was then used to evaluate the same threshold-based
figure of merit as in the main text.\\
\Cref{fig:S5}a--c summarize this thin-film optimization. The selected thicknesses are
$d_1=\SI{320}{nm}$ for the upper TiO$_2$ layer and $d_2=\SI{910}{nm}$ for the lower
a-Si layer. The reflectance map contains a continuous low-reflection locus that spans
the effective-index range required for the lower a-Si phase library. Along this locus,
the reflected contributions from the effective interfaces destructively interfere, as in
the same-material TiO$_2$/TiO$_2$ design.\\
The effective-index targets are then translated into physical nanopillar geometries.
In the hybrid case, the geometry-to-index relation is material dependent. We therefore
calculate separately the fundamental-mode effective indices of periodic square TiO$_2$
and a-Si nanopillars with pitch $U=\SI{450}{nm}$, as shown in \cref{fig:S5}d. Let
$g_{\mathrm{aSi}}(w_2)$ map the lower a-Si side width to its effective index and
$g_{\mathrm{TiO_2}}(w_1)$ map the upper TiO$_2$ side width to its effective index. The
hybrid bilayer design rule is then
\begin{equation}
w_1 =
g_{\mathrm{TiO_2}}^{-1}
\left[
f\!\left(g_{\mathrm{aSi}}(w_2)\right)
\right].
\label{eq:S_hybrid_design_rule}
\end{equation}
Thus, for each lower a-Si nanopillar chosen to provide the required transmission phase,
the upper TiO$_2$ nanopillar is selected to satisfy the antireflection mapping.\\
\Cref{fig:S5}e compares the simulated phase and transmittance of a conventional
single-layer a-Si library with those of the hybrid TiO$_2$/a-Si bilayer library
(\cref{fig:S5}f). Both libraries provide full $0$ to $2\pi$ phase coverage, but the
single-layer a-Si library exhibits a pronounced transmittance drop for large side widths
because of the strong effective-index contrast between the nanopillar mode and the
surrounding media. The hybrid bilayer substantially suppresses this loss. The minimum
transmittance increases from $56.5\%$ for the single-layer a-Si library to $94.7\%$ for
the TiO$_2$/a-Si bilayer library, while the mean transmittance increases from $90.6\%$
to $97.8\%$. These results confirm that the bilayer antireflection strategy can be
extended beyond same-material platforms and can be combined with high-index
materials to improve both phase sampling and optical throughput.

\clearpage

\section{Loss decomposition and substrate-normalized focusing efficiencies}
\label{sec:loss_channels}

To identify the physical origin of the efficiency enhancement in the bilayer metalens,
we performed full-device finite-difference time-domain (FDTD) simulations for three
configurations: a bare glass substrate, a conventional single-layer TiO$_2$ metalens on
glass and an antireflective TiO$_2$/TiO$_2$ bilayer metalens on glass. In all cases, an
$x$-polarized plane wave was incident from air, and the simulations included the full
metalens aperture rather than a locally periodic or phase-only approximation \cite{hughes2021}.

\begin{figure}[h]
\centering
\suppfigure{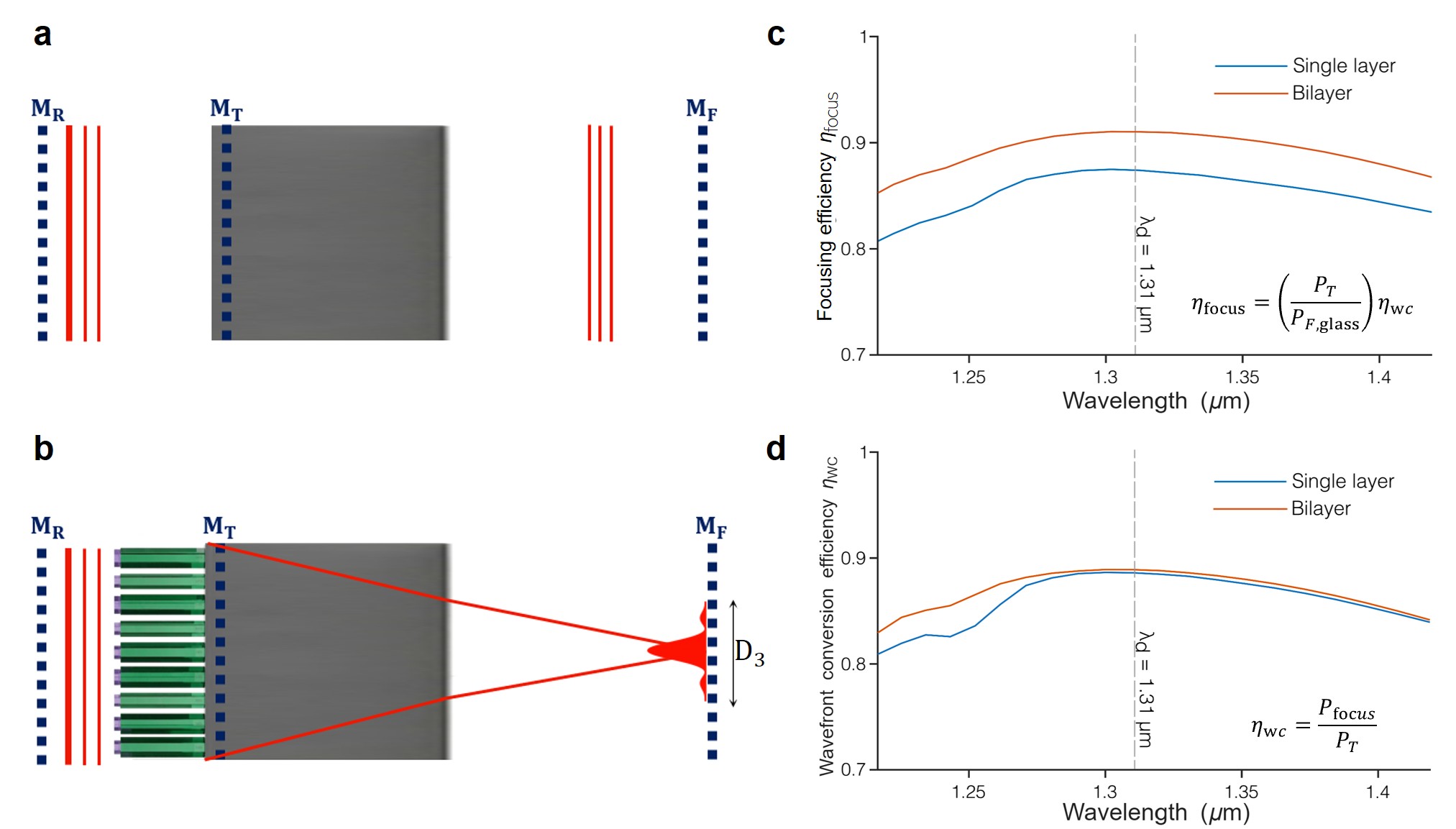}{0.99\linewidth}
\caption{\textbf{Separation of reflection loss and wavefront-conversion efficiency.}
\textbf{a,} Bare-glass reference simulation. Monitor \(M_F\) is placed at the output
plane and records the transmitted power through the unpatterned glass substrate,
\(P_{F,\mathrm{glass}}\).
\textbf{b,} Full-device metalens simulation. Monitor \(M_R\) records the
backward-propagating reflected power \(P_R\), monitor \(M_T\) records the
forward-propagating power \(P_T\) transmitted into the substrate immediately after the
metasurface, and monitor \(M_F\) is placed at the output/focal plane. The focal power
\(P_{\mathrm{focus}}\) is obtained by integrating the intensity at \(M_F\) over a circular
region centred on the focus whose radius is set by the third zero of the corresponding
diffraction-limited Airy pattern.
\textbf{c,} Substrate-normalized focusing efficiency
\(\eta_{\mathrm{focus}}=P_{\mathrm{focus}}/P_{F,\mathrm{glass}}\). The higher value for the
bilayer metalens indicates that more optical power is delivered to the designed focal
region.
\textbf{d,} Wavefront conversion efficiency
\(\eta_{\mathrm{wc}}=P_{\mathrm{focus}}/P_T\) for the single-layer and bilayer
TiO\(_2\)/TiO\(_2\) metalenses. The nearly identical values indicate that the two devices
convert the transmitted field into the designed focus with comparable fidelity.}
\label{fig:S4} 
\end{figure}
\noindent
The monitor configuration is shown schematically in \cref{fig:S4}a,b. Monitor $M_R$ is
placed in the incident medium (air) and records the backward-propagating power $P_R$, from
which the full-device reflectance is obtained as
\begin{equation}
R_{\mathrm{dev}}=\frac{P_R}{P_{\mathrm{inc}}},
\label{eq:S_device_reflectance}
\end{equation}
where $P_{\mathrm{inc}}$ is the incident power. Monitor \(M_T\) is placed inside the glass substrate immediately after the metasurface
and records the forward-transmitted complex field and the corresponding transmitted
power \(P_T\). The output/focal plane \(M_F\) is obtained in post-processing by propagating the
recorded complex field using the angular-spectrum method, i.e. an angular-spectrum
implementation of the Rayleigh--Sommerfeld diffraction integral ~\cite{Matsushima2009}.\\
In this approach, the complex field recorded at $M_T$ is decomposed into transverse spatial-frequency
components and propagated using the angular-spectrum method. Within the substrate,
propagation is described by the phase factor $\exp(i k_{z,g} z_g)$, where $k_{z,g}$
is the longitudinal wave-vector component in glass. If desired, the same propagated
field can be continued into air by applying the angle-dependent Fresnel transmission
coefficients at the glass--air interface before further free-space propagation.

\noindent
For bare-glass reference simulation, the total forward power evaluated at this
propagated plane is denoted \(P_{F,\mathrm{glass}}\). For the metalens simulations, the
focal power \(P_{\mathrm{focus}}\) is obtained by integrating the propagated intensity at
\(M_F\) over a circular region centred on the designed focus and extending to the third
zero of the corresponding diffraction-limited Airy pattern.
The integration radius is
$\rho_3=(u_3/2\pi)(\lambda/\mathrm{NA})$, where $u_3=10.1735$ is the third zero of
$J_1(u)$; equivalently, the integration diameter is
$D_3\simeq 3.24\lambda/\mathrm{NA}$. The resulting power is denoted
$P_{\mathrm{focus}}$.
The focusing efficiency reported in the main text is defined as
\begin{equation}
\eta_{\mathrm{focus}}=
\frac{P_{\mathrm{focus}}}{P_{F,\mathrm{glass}}},
\label{eq:S_focusing_efficiency}
\end{equation}
which normalizes the power delivered to the designed focal region by the power
transmitted through the same unpatterned glass substrate. This substrate-normalized
definition matches the experimental normalization while retaining all loss channels
present in the complete device simulation.\\
However, $\eta_{\mathrm{focus}}$ alone does not determine whether an efficiency difference
originates from reduced reflection or from a change in wavefront formation. Power can
be lost from the focal region through phase errors, amplitude nonuniformity,
finite-pitch discretization, scattering into unwanted spatial components and residual
side lobes. To separate optical throughput from wavefront-conversion quality, we define
the wavefront conversion efficiency
\begin{equation}
\eta_{\mathrm{wc}}=
\frac{P_{\mathrm{focus}}}{P_T}.
\label{eq:S_wavefront_efficiency}
\end{equation}
This metric gives the fraction of power already transmitted into the substrate that is
subsequently delivered to the designed focal region. It is therefore largely insensitive
to input-side reflection loss and instead probes how efficiently the transmitted field is
converted into the target focus.
Combining Eqs.~\eqref{eq:S_focusing_efficiency} and
\eqref{eq:S_wavefront_efficiency} gives
\begin{equation}
\eta_{\mathrm{focus}}=
\left(\frac{P_T}{P_{F,\mathrm{glass}}}\right)\eta_{\mathrm{wc}},
\label{eq:S_efficiency_decomposition}
\end{equation}
where the first factor is a substrate-normalized throughput term and the second factor
describes wavefront conversion. This decomposition distinguishes two possible origins
of an efficiency enhancement. If the single-layer and bilayer metalenses have comparable
$\eta_{\mathrm{wc}}$, point-spread functions and Strehl ratios, then they convert the
transmitted field into the focus with comparable fidelity. In that case, a difference in
$\eta_{\mathrm{focus}}$ is primarily attributable to the throughput term, and therefore to the
amount of optical power coupled into the substrate.\\
This is the case for the TiO$_2$/TiO$_2$ metalenses studied in the main text. Around
the design wavelength, the bilayer device shows a larger substrate-normalized focusing
efficiency than the single-layer reference (\cref{fig:S4}c), while the two devices exhibit
nearly identical wavefront conversion efficiencies (\cref{fig:S4}d). At \(\lambda=\SI{1310}{nm}\), the wavefront-conversion efficiency is
\(\eta_{\mathrm{wc}}=0.886\) for the single-layer metalens and
\(\eta_{\mathrm{wc}}=0.889\) for the bilayer metalens, differing by only
\(0.30\) percentage points. Here, the focusing efficiency is evaluated for the field
propagating inside the glass substrate and is normalized by the power transmitted into
the same unpatterned glass substrate.

\noindent
By contrast, the simulated full-device reflectance decreases from 5.6\% for
the single-layer metalens to 2.0\% for the bilayer metalens, corresponding to
a reduction of 3.6 percentage points. This reduction is accompanied by an
increase in the into-substrate, substrate-normalized focusing efficiency from
87.4\% to 91.1\%. Because the
wavefront-conversion efficiencies differ only minimally whereas the reflectance decreases
substantially, the bilayer efficiency enhancement is dominated by increased optical
throughput into the substrate rather than by improved wavefront formation.

\clearpage

\section{Hybrid TiO$_2$/a-Si bilayer metalens performance}
\label{sec:hybrid_metalens}

To verify that the hybrid-library improvement translates to a complete wavefront-shaping
device, we designed two metalenses: one based on the conventional single-layer a-Si
library and one based on the antireflective TiO$_2$/a-Si bilayer library. The purpose of
this comparison is not to optimize the absolute performance of the hybrid platform
against the TiO$_2$/TiO$_2$ device discussed in the main text, but to test whether the
same antireflective design principle remains effective under different operating
conditions: a different wavelength, a different material platform and a more demanding
wavefront. We therefore design the hybrid devices at $\lambda_d=\SI{1550}{nm}$ with
numerical aperture $\mathrm{NA}=0.36$. Both devices use the same hyperbolic phase
profile, diameter $D=\SI{50}{\micro m}$ and focal length $f=\SI{100}{\micro m}$ into
the glass substrate. The full devices were simulated by FDTD under normally incident
$x$-polarized illumination.

\begin{figure}[h]
\centering
\suppfigure{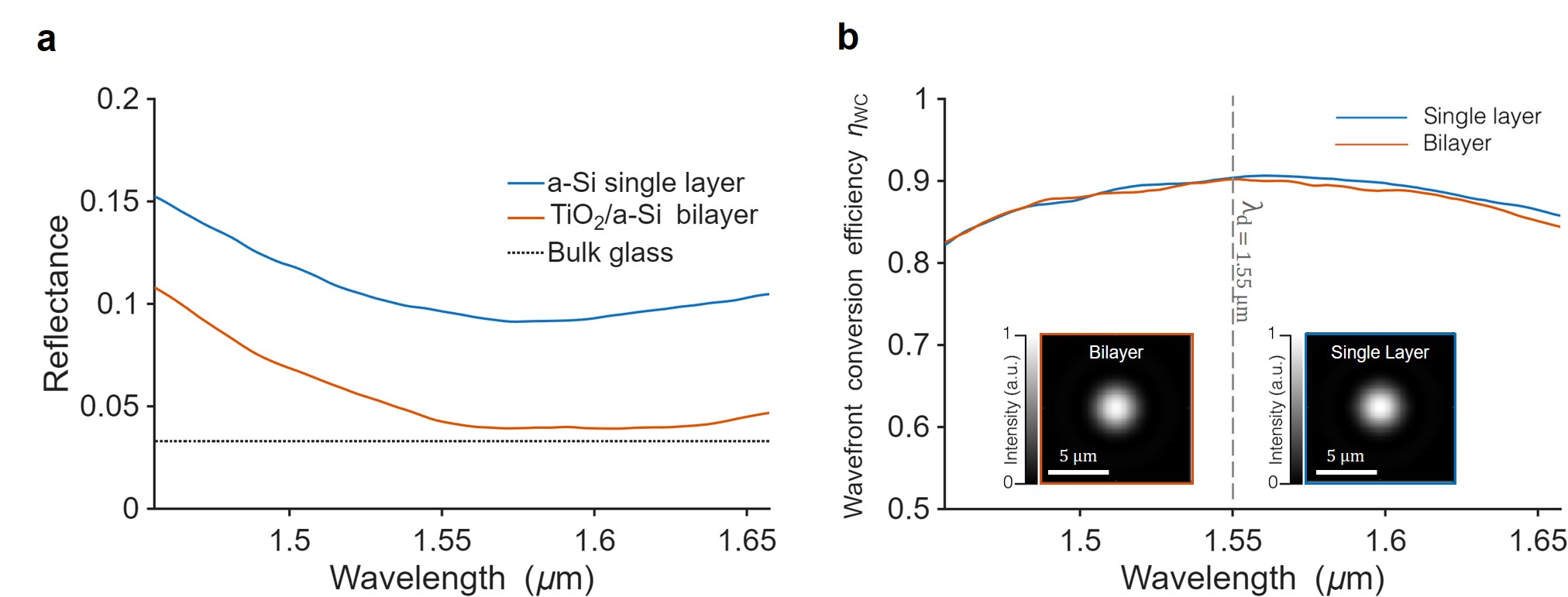}{0.85\linewidth}
\caption{\textbf{Simulated performance of a single-layer a-Si metalens and an
antireflective TiO$_2$/a-Si hybrid bilayer metalens.}
\textbf{a,} Reflectance spectra of the single-layer a-Si metalens, the TiO$_2$/a-Si bilayer
metalens and a bare glass substrate around the design wavelength
$\lambda_d=\SI{1550}{nm}$. At the design wavelength, the reflectance decreases from
$R_{\mathrm{dev}}=0.096$ for the single-layer a-Si metalens to
$R_{\mathrm{dev}}=0.043$ for the hybrid bilayer metalens.
\textbf{b,} Wavefront-conversion efficiency of the two metalenses. At the design wavelength,
$\eta_{\mathrm{wc}}=0.902$ for the single-layer a-Si metalens and
$\eta_{\mathrm{wc}}=0.904$ for the TiO$_2$/a-Si bilayer metalens, indicating comparable
wavefront-conversion quality. Insets show the corresponding simulated point-spread
functions at the design wavelength. Scale bar, $\SI{2}{\micro m}$.}
\label{fig:S6}
\end{figure}
\noindent
The simulated reflectance spectra are shown in \cref{fig:S6}a, together with the
reflectance of a bare glass substrate. At the design wavelength, the single-layer a-Si
metalens reflects $R_{\mathrm{single}}=0.096$ of the incident power, whereas the
TiO$_2$/a-Si bilayer metalens reflects $R_{\mathrm{bilayer}}=0.043$. Thus, the hybrid
bilayer reduces the device-level reflectance by $5.3$ percentage points, corresponding
to a relative reduction of approximately $55\%$ compared with the single-layer a-Si
reference. The absolute reflectance remains larger than in the TiO$_2$/TiO$_2$ metalens
studied in the main text, as expected for a higher-index material platform and a different
operating wavelength. The relevant comparison is therefore between the hybrid bilayer
and its single-layer a-Si control, which shows that the impedance-matching strategy
remains effective beyond the same-material TiO$_2$/TiO$_2$ system.\\
The transmitted near fields were then propagated to the focal plane and used to
calculate the point-spread functions. The wavefront-conversion efficiencies, obtained by
integrating the focal intensity over the aperture defined in \cref{sec:loss_channels}, are
shown in \cref{fig:S6}b. At the design wavelength, the wavefront-conversion efficiency
is \(\eta_{\mathrm{wc}}=0.902\) for the single-layer a-Si metalens and
\(\eta_{\mathrm{wc}}=0.904\) for the TiO\(_2\)/a-Si bilayer metalens. The nearly identical
values, together with the similar point-spread functions, indicate that the two devices
convert the transmitted field into the designed focus with comparable fidelity.\\
The into-substrate focusing efficiency can be estimated using the same decomposition
introduced in \cref{sec:loss_channels},
\begin{equation}
\eta_{\mathrm{focus}}
=
\left(
\frac{P_T}{P_{T,\mathrm{glass}}}
\right)
\eta_{\mathrm{wc}}
\simeq
\left(
\frac{1-R_{\mathrm{dev}}}{1-R_{\mathrm{glass}}}
\right)
\eta_{\mathrm{wc}},
\label{eq:S_hybrid_efficiency_decomposition}
\end{equation}
where \(P_{T,\mathrm{glass}}\) is the power transmitted into the unpatterned glass
substrate and \(R_{\mathrm{glass}}\) is the single-interface Fresnel reflectance of the
bare glass substrate. 
\noindent
Using \(R_{\mathrm{glass}}\simeq0.033\), at
\(\lambda_d=\SI{1550}{nm}\) the simulated device reflectance is
\(R_{\mathrm{dev}}=0.096\) for the single-layer a-Si metalens and
\(R_{\mathrm{dev}}=0.043\) for the TiO\(_2\)/a-Si bilayer metalens.
The corresponding substrate-normalized throughput factor,
\((1-R_{\mathrm{dev}})/(1-R_{\mathrm{glass}})\), is therefore approximately
\(0.934\) for the single-layer a-Si metalens and \(0.990\) for the
TiO\(_2\)/a-Si bilayer metalens. Combining these throughput factors with
the simulated wavefront-conversion efficiencies,
\(\eta_{\mathrm{wc}}=0.902\) for the single-layer metalens and
\(\eta_{\mathrm{wc}}=0.904\) for the bilayer metalens, gives
\(\eta_{\mathrm{focus}}\simeq84.2\%\) for the single-layer a-Si metalens and
\(\eta_{\mathrm{focus}}\simeq89.5\%\) for the hybrid bilayer metalens.\\
This result also illustrates the complementary roles of wavefront sampling and
antireflection. Although the TiO\(_2\)/a-Si bilayer has a higher device
reflectance than the TiO\(_2\)/TiO\(_2\) bilayer platform, the higher refractive
index of a-Si allows the required \(0\) to \(2\pi\) phase coverage to be achieved
with a smaller unit cell, \(U=\SI{450}{nm}\), compared with
\(U=\SI{600}{nm}\) for the TiO\(_2\)/TiO\(_2\) design. This finer spatial
sampling reduces phase-discretization and wavefront-conversion errors at
\(\mathrm{NA}=0.36\). For comparison, a TiO\(_2\)/TiO\(_2\) design at the same
numerical aperture reaches a simulated focusing efficiency of approximately
\(77.0\%\).\\
The hybrid simulation therefore highlights a broader design opportunity: high-efficiency
metasurfaces can benefit from the simultaneous reduction of wavefront-conversion
errors, enabled here by the smaller lattice period, and reflection losses, enabled by the
bilayer antireflection strategy. These results confirm that the bilayer antireflection
framework can be extended across wavelengths, material systems and device numerical
apertures.

\clearpage

\section{Interlayer-alignment sensitivity of the TiO$_2$/TiO$_2$ bilayer metalens}
\label{sec:alignment}

Interlayer alignment is an important fabrication consideration for bilayer metasurfaces
because the antireflective response depends on the relative registration of the two
patterned layers. A global lateral displacement of the complete bilayer stack with
respect to the substrate does not change the optical response, since the glass substrate
is laterally homogeneous. The relevant fabrication error is therefore the relative lateral
displacement between the upper nanopillar layer, adjacent to air, and the lower
nanopillar layer, adjacent to the substrate.

\begin{figure}[h]
\centering
\suppfigure{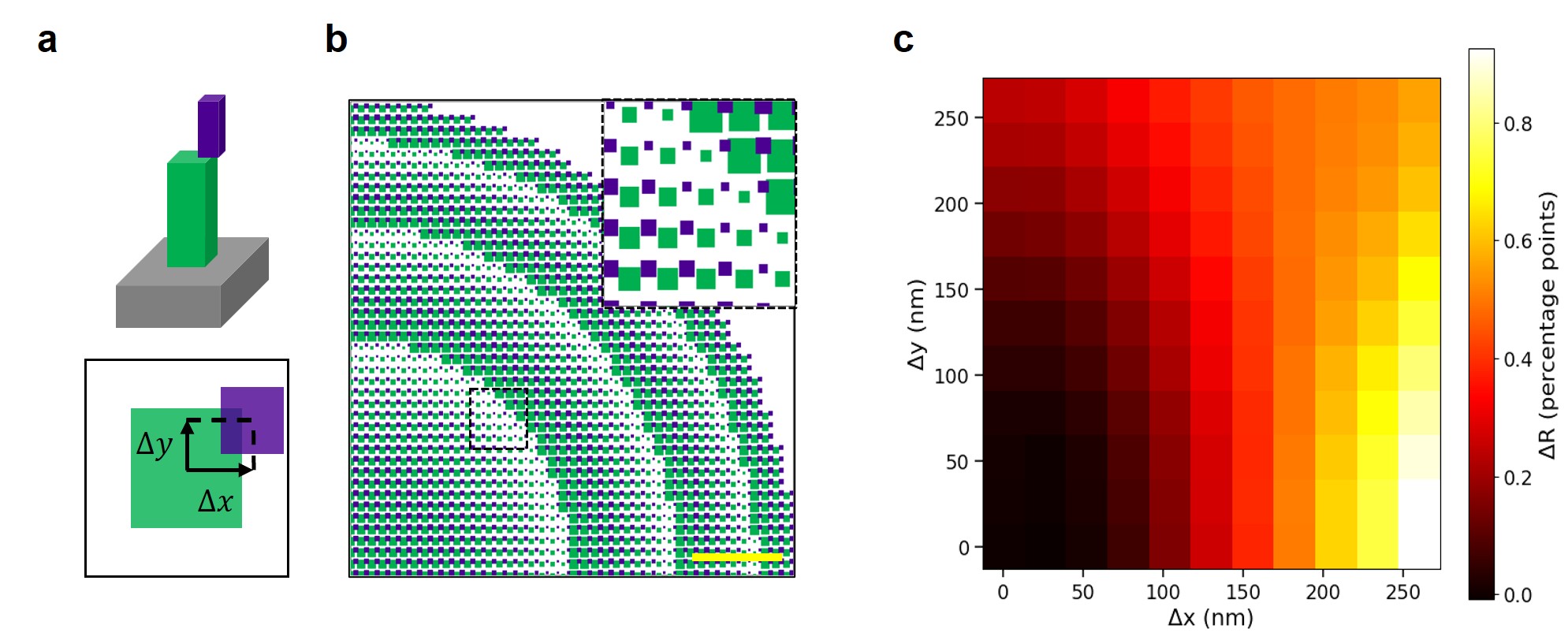}{0.95\linewidth}
\caption{\textbf{Interlayer-alignment sensitivity of the TiO$_2$/TiO$_2$ bilayer metalens.}
\textbf{a,} Definition of the lateral interlayer displacement $(\Delta x,\Delta y)$ between
the upper and lower TiO$_2$ nanopillars.
\textbf{b,} Representative section of a bilayer metalens with a uniform lateral displacement
applied to the upper layer while the lower layer is kept fixed.
\textbf{c,} Simulated reflectance change $\Delta R = R(\Delta x, \Delta y) - R(0, 0)$ of the full bilayer device at $\lambda = \SI{1310}{nm}$ as a function of interlayer displacement, under normally incident $x$-polarized illumination. Displacements are swept from $0$ to $\SI{260}{nm}$ along both in-plane axes; the negative quadrants follow by the $C_{4v}$ symmetry of the design. The expected interlayer alignment precision of the electron-beam lithography tool ($\approx\!\SI{20}{nm}$) lies well within the region where alignment-induced reflection is negligible ($\Delta R < 0.1\%$).}
\label{fig:S7}
\end{figure}

\Cref{fig:S7}a defines the lateral interlayer displacement $(\Delta x,\Delta y)$ at the
meta-atom level. In the simulations, this displacement was applied uniformly to the
entire upper layer while the lower layer was kept fixed. A representative section of a
misaligned metalens is shown in \cref{fig:S7}b. For each imposed displacement, the
complete device was simulated by FDTD under normally incident $x$-polarized
illumination, and the reflected power was recorded to obtain the full-device reflectance.

\Cref{fig:S7}c shows the simulated reflectance at $\lambda=\SI{1310}{nm}$ as a function
of interlayer displacement over one approximately half a lattice period along each in-plane direction. The
reflectance remains below that of bare glass over a broad range of offsets, indicating
that the antireflective response is tolerant to moderate overlay errors. This tolerance is
well matched to the fabrication process used here: the electron-beam lithography tool
provides an expected interlayer alignment precision of approximately $\SI{20}{nm}$,
which is much smaller than the displacement scale over which the simulated reflectance
changes appreciably.

The response is not symmetric under exchange of $\Delta x$ and $\Delta y$ for the
fixed $x$-polarized excitation. This asymmetry arises because lateral misalignment
breaks the in-plane four-fold symmetry of the aligned square bilayer meta-atoms,
making offsets parallel and perpendicular to the incident polarization optically
inequivalent.

The largest reflectance increase occurs for intermediate offsets rather than for a
full-period displacement. This trend is consistent with a spatial-overlap picture of the
bilayer antireflection mechanism. When the upper pillar is displaced by a fraction of
the lattice period, its overlap with the corresponding lower pillar is reduced. A larger
fraction of the incident field then interacts directly with the lower high-index pillar,
increasing the effective index contrast at the entrance side of the structure. At the same
time, reflected fields generated near the lower facet of the upper pillar are less
effectively matched by the lower layer. Both effects weaken the destructive interference
between the reflected contributions from the effective interfaces and increase the total
reflectance.

We note that some large-offset configurations considered here may be mechanically
unrealistic, because portions of the upper pillars would be weakly supported or
effectively floating. The map in \cref{fig:S7}c should therefore be interpreted as an
optical sensitivity analysis of idealized interlayer displacements, rather than as a
complete mechanical-yield model.